\RequirePackage{fix-cm}
\documentclass[twocolumn]{svjour3}          % twocolumn
\smartqed  % flush right qed marks, e.g. at end of proof
\usepackage{graphicx}
\usepackage{color}
\usepackage{booktabs}

\usepackage{lineno}
\usepackage[round]{natbib}
 \journalname{\textsc{The Astrophysical Journal}}

\begin{document}

\title{
Dense Particle Clouds in Laboratory Experiments in Context of Drafting and Streaming Instability 
}

%
% ^.
\subtitle{Studying the Transition from Single Particle to Collective Particle Behavior in the Knudsen regime.}%the transition regime from Knudsen to Stokes drag.}

\titlerunning{Dense Particle Clouds in Laboratory Experiments}        % if too long for running head

\author{Niclas Schneider$^1$, Gerhard Wurm$^1$, Jens Teiser$^1$, Hubert Klahr$^2$, Vincent Carpenter$^2$}


\institute{$^1$             University of Duisburg-Essen, Lotharstr. 1-21, 47057 Duisburg, Germany\\
              Tel.: +49-(0)203-379-1641\\
              Fax: +49-(0)203-379-1965\\
              \email{niclas.schneider@uni-due.de}\\                   %  \\
           \and 
           $^2$   Max Planck Institute for Astronomy, K{\"o}nigstuhl 17, 69117 Heidelberg, Germany\\
        Tel.:  +49-(0)6221-528-255\\
        \email{klahr@mpia.de}
}

\date{Received: 2018 June 15 / Accepted: 2019 January 8 / Published: 2019 February 6}

\maketitle

\begin{abstract}

{
The streaming instability, as an example of instabilities driven by particle feedback on a gas flow, 
has been proven to have a major role in controlling the formation
of planetesimals. 
%Collective feedback via friction from particles to the gas flow in which they are embedded is known to trigger instabilities and has been proven to have a major role in controlling the formation of planetesimals. These instabilities in protoplanetary disks occur at the transition from being gas dominated to being dust and ice particle dominated. 
Here, we present experiments to approach this situation in the laboratory for particles in the Knudsen flow regime.
In these experiments, we observe a particle cloud trapped for about 30~s in a rotating system under Earth's gravity.
For average dust-to-gas ratios up to 0.08, particles behave like individual test particles. Their sedimentation speed is identical to that of a single free-falling particle, even in locally denser regions.  
However, for higher dust-to-gas ratios, the motion of particles becomes sensitive to clumping. Particles in locally denser regions now sediment faster. Their sedimentation speed then depends {linearly} on the overall dust-to-gas ratio. This clearly shows a transition from tracerlike behavior to collective behavior. 
Beyond {these} findings, these types of experiments 
can now be used as a gauge to test 
the particle feedback models in astrophysical hydrocodes which are currently used for numerical simulations of streaming instabilities.}

\end{abstract}

\section{Introduction}
In the core accretion model of planet formation, dust first has to be converted to kilometer-sized planetesimals, which then accumulate via gravitational interaction into planetary embryos, the precursors of terrestrial planets as well as of gas and ice giants.
In the standard scenario, terrestrial planet formation starts from dust, which evolves through all size scales. Initially, dust growth is driven by mutual collisions \citep{blumwurm2008}. However, this growth gets stalled either at the {drift barrier \citep{Weidenschilling1977a}, the fragmentation barrier \citep{Birnstiel2012}} or the bouncing barrier \citep{zsom2010,Kelling2014,kruss2016,kruss2017}.

One way to continue evolution is the gravitational collapse of the particle layer in the disk \citep{Safronov1969,Goldreich1973}. Dense particle layers with local dust-to-gas ratios of unity and above are subject to instabilities \citep{Weidenschilling1980} driven by the mutual coupling between gas and particles, which have an incommensurable equilibrium state \citep{1986Icar...67..375N}. These instabilities can be generalized as resonant drag instabilities \citep{Squire2018SI}, of which the streaming instability plays a major role in regulating the onset of gravitational collapse \citep{Youdin2005, Johansen2015, simon2016, schreiber2018}. Only when particles are concentrated to the point where self-gravity starts to dominate over turbulent diffusion can kilometer-size planetesimals form by gravitational collapse \citep{Johansen2007, 2011A&A...529A..62J, Johansen2015}.

In general the numerical simulations show that high solid-to-gas ratios are required and that the gas-grain coupling times have to be comparable to the orbital timescales. For example, \citet{bai2010} found the typical dust-to-gas ratio of 0.01 to be insufficient to concentrate solids to the limit of gravitational collapse.
In any case, the scenario of {gravitational collapse regulated by} streaming instabilities is widely accepted as the standard way to connect to and take over from the preceding collisional growth phases of pebbles. However, laboratory experiments are essentially missing. 

{The situation of a protoplanetary disk can certainly not be set up 1 to 1 in a laboratory setting. However, the motion of a dense cloud of particles over a longer timescale can be accessed as described in this paper. It might answer basic questions, like under what conditions are particles only tracers of gas motion or when is a back-reaction to the gas notable? Does this really occur only at a dust-to-gas ratio of about 1? {What are the minimum dust-to-gas ratios needed to see collective effects? Is there a smooth or a sharp transition? How do density fluctuations develop in experiments? Are they stable or do they grow in amplitude?} As numerical codes describing large protoplanetary disks should be capable of reproducing small-scale laboratory results, this also offers an opportunity to verify and improve numerical codes. This study is a first work in that direction.}

{So far, only} \citet{Lambrechts2016} have hinted at a potential experiment to study particle density fluctuations in the laboratory for sedimenting grains, which they simulated in a static gas column. A related experiment was carried out by \citet{Capelo2018}. That work studied sedimenting grains in the upward draft of a gas flow on a microscopic level. 
{According to  \citet{Capelo2018}, the sedimentation velocities of particle pairs are affected within a vertical interparticle distance of 4~mm. {However, \cite{Capelo2018} only observed a small volume of 1 $\rm cm^3$ of a meter-long 9 cm diameter tube, making it impossible to study large-scale influences,} and they generally observed low absolute particle numbers $\ll 50$. Also, the grains in their experiment had an unknown size distribution, and there is a potential influence on continuous wall collisions in various ways, i.e. introducing inhomogeneities could have been introduced during redispersion. 
%Therefore, up to 650 particles in a larger volume covering the whole experimental setup are observed to get a better insight in high particle loading.
}
%{but due to generally low dust-to-gas ratios no collective behavior of a larger number of grains could be observed. } 
 
However, no other experimental approaches to streaming in low-pressure gases are known to the authors.
 Having a complex two-fluid (particles and gas) problem, it is of greatest interest to
us to study in an experiment how particle clouds really react {to changing dust-to-gas ratios.}

{Usually, {$\rm 100 \mu m$ particles fall to the bottom of a decimeter-sized vessel at low pressure on a timescale of only 1~s. Here, we are interested in long-time behavior and interparticle influences to investigate possible particle concentration mechanisms over time as predicted by streaming or drag instabilities.} Therefore, ways to levitate grains are needed, such as the upward draft used by \citet{Capelo2018}. We use a different approach here. An experimental setup that allows the study of particle evolutions for {a} longer time in the laboratory in a more defined way is a rotating cylinder with a cloud of particles and gas within. The rotation axis has to be horizontal. In such a rotating cylinder with no-slip boundaries at the cylinder mantle, the gas motion in equilibrium is simply a rigid rotation together with the cylinder. Solid particles embedded in this cylinder still undergo sedimentation. However, if their gas-grain friction time is short compared to the rotation time, they can be considered to fall with terminal speed at all times. This terminal speed is always given relative to the gas. As the gas motion depends on the distance from the cylinder axis, so does the absolute motion of the grain in the laboratory reference frame.}
In the updrift part of the gas rotation, there is a stable point for a sedimenting particle, where its downward sedimentation speed and the upward gas velocity cancel each other.
{Disturbing a particle from this equilibrium point leads to circular trajectories as shown below. This way particles can be trapped for {a} long time, mostly limited by centrifugal forces as also estimated below.}

Such an experiment has been used before by \citet{blum1998} and \citet{PoppeBlum1997}, but for aggregation studies. This setup is inspired by the idea of particle trapping in convective eddies in protoplanetary disks as studied by \citet{klahr1997}. Details are given in the experiment section.

In a simplified way, the situation that can be studied in the laboratory with the given setup is as follows. A larger grain takes longer to couple to a gas flow and 
sediments faster than a smaller grain.
But how many small particles have to be how close to each other to act like a larger grain, {as suggested by \citet{JohansenYoudin2007}}? 
And how does increasing the overall dust-to-gas ratio influence the particle motion in this 
%rotating 
system? {Will "drafting", the attraction of particles in the wake of a particle clump, as outlined by \citet{Lambrechts2016}, change the particle density and spontaneously form dense clumps?}

{In section 2 we outline the basic gas-grain interactions and parameters describing clouds of particles especially in the experiments. We also introduce closeness as a parameter. Section 3 describes the experiment, its function, and its basic parameters. Section 4 gives the results of measurements of individual particles moving as part of a particle cloud. Section 5 adds some more discussion to the results. Section 6 is a short summary of the most important experimental findings. Being far from perfect, we point out caveats in section 7 and conclude in section 8.}

\section{Grain-Gas Interaction}

The motion of a single particle embedded in a gas depends on the gas-grain coupling time or friction time $\tau_f$. 

%It can be defined as

%\begin{equation}
%\tau_f = \frac{m}{F} \cdot v = a \cdot v
%\end{equation}

%where $F$ is the drag force acting on the particle of mass $m$ which moves relative to the gas with a speed $v$. The force accelerates the particle with acceleration $a$. 

$\tau_f$ is only well defined (constant) in flows where the drag force is proportional to the gas speed. This is valid for small Reynolds numbers, which is the case here {(see table \ref{para})}. In particular, the coupling time depends on the flow regime, which can be described by the Knudsen number $Kn$.
The Knudsen number is the ratio between the mean free path length between the molecules $\lambda$ and the particle radius $r$,
\begin{equation}
Kn = \frac{\lambda}{r}
\label{tau_no}
\end{equation}
For small $Kn\ll1$ the flow is continuous. For large $Kn\gg1$ the flow is molecular.
For molecular flow, the coupling time for a spherical particle of radius $r$ is given as
\begin{equation}
\tau_{f_E} = \frac{4}{3} \frac{\rho_p}{\rho_g} \frac{r}{v_g}
\label{tau0}
\end{equation}
with $\rho_g$ as the gas density, $\rho_p$ as the particle density and $v_g$ as the thermal gas velocity. 
For small $Kn$, the Stokes law applies and it is
\begin{equation}
\tau_{f_{\rm S}} = \frac{2 r^2 \rho_p}{9 \eta} 
\label{tauS}
\end{equation}
with viscosity $\eta$.
For $Kn \sim 1$ eq. \ref{tauS} can be used, if a correction factor $f_c$ is added:
\begin{equation}
\tau_f = \tau_{f_{\rm S}} \cdot f_c.
\label{tau}
\end{equation}
Here, $f_c$ is the Cunningham correction \citep{cunningham1910,hutchins1995}:
\begin{equation}
f_c = 1+Kn(1.257+0.4e^{-0.55 Kn})
\end{equation}
If a particle is dragged through a fluid with constant external force $F_{\rm ext}$ it will be accelerated until the drag compensates for this force.
This leads to a constant velocity $v_{\rm rel}$ relative to the gas in equilibrium. In the case of gravity with gravitational acceleration $g$ it is
\begin{equation}
v_{\rm rel} = g \cdot \tau_f
\label{vrel}
\end{equation}
However, if the gas motion changes on a timescale $\tau_{\rm gas}$ comparable to the friction time, this simplification no longer holds. Therefore, an important quantity for describing the system is the Stokes number, which is the ratio between coupling time and typical time for gas motion variation:
\begin{equation}
St_0 = \frac{\tau_f}{\tau_{\rm gas}.}
\end{equation}
{In turbulence, $\tau_{\rm gas}$ usually describes the correlation time of the smallest turbulent eddies in the flow.
A different $\tau_{\rm gas}$ plays a major role in our experiments, where we take $\tau_{\rm gas}$ to be the rotational timescale $\tau_{\rm gas} = 1/(2\pi f) = 1/\Omega$ of the experiment chamber, as the gas follows this motion. Here, $f$ is the rotation frequency and $St = \tau_f \cdot 2\pi~f$. 

This Stokes number, for example, allows an estimate of the time that a particle can stay trapped in the rotating flow or of the timescale on which the cloud gets thinner, limited by centrifugal losses as follows.
The radial drift velocity resulting from the centrifugal acceleration from the rotating chamber is
\begin{equation}
v_r = \tau_f \Omega^2 r.
\end{equation}
To first order, the time $t_d$ needed to double this orbit is determined by $v_r t_d = r$. As $t_d$ is given by the number of rotations $N_d$ and the rotation frequency by $t_d = N_d / f$, we can estimate the number of rotations $N_d$ to double the orbits by \citep{klahr1997}
\begin{equation}
N_d = \frac{1}{2 \pi St}.
\end{equation}
With our chosen $St = 0.014$ for the experimental setup (see below), typical  changes, for example, of the dust-to-gas ratio occur in about 10 rounds or 30 s.

The Stokes number as referred to in the planet formation context $St_{\rm Kepler} = \tau_f \Omega_{\rm Kepler}$ has a different meaning. 
Here, one uses $\Omega_{\rm Kepler}$, the orbital frequency of the disk.
This Stokes number scales the radial drift rate of particles in the nebula to \citep{Weidenschilling1977a}
\begin{equation}
v_r \propto - 2 \frac{St_{\rm Kepler}}{1 + St^2_{\rm Kepler}},
\end{equation}
indicating that a maximum of the radial inward drift occurs for $St_{Kepler} = 1$.

From work on streaming in\-sta\-bi\-li\-ties, significant \linebreak growth rates are ex\-pected to start for $St_{\rm Kepler} \geq 0.001$ -- 
$St_{\rm Kepler} \geq 0.01$ \citep{bai2010, drkazkowska2014, yang2017, Carrera2017},{ because drift {between two components} is essential for the onset of any resonant drag instability.

In both cases, experiment and disk, a low Stokes number indicates that the mean motion of particles, as either radial drift toward the star or sedimentation and centrifugal motion in the laboratory, can be described by a terminal velocity to first order (eq. \ref{vrel}).}
Eq. \ref{vrel} holds strictly speaking only for small Stokes numbers $St_0$, which should be given in our low Reynolds number (${\rm Re} = \frac{\rho_{\rm gas} \cdot v \cdot d}{\eta} = 0.007$ for a single sphere) quasilaminar setup of
rotating gas in the experimental chamber.
In other words $St_0$ will initially be smaller than $St$, but this may change if the particles themselves drive additional gas velocities. A violation of eq. \ref{vrel} then is the route to further increase the local dust-to-gas ratio via turbulence (independently of whether we are considering the solar nebula or a laboratory flow), which otherwise may be hard to achieve \citep{Johansen2005}. Maximum concentration in a turbulent flow occurs for $St_0=1$ particles \citep{Ormel2007}.
While the linear streaming instability can indeed concentrate radially drifting particles in the solar nebula  \citep{Youdin2005}, such a linear instability, so far, is not known for sedimenting spherical particles in the laboratory. For nonspherical grains, an instability could be identified \citep{koch_shaqfeh_1989}}. 
{It would be interesting to study the differences between spherical and nonspherical grains. As the given experiment and analysis is already complex and is used for the first time, we start with spherical particles. Given their complexity, the choice and acquiring nonspherical particles including all aspects of describing and manipulating nonsphericity, rotation, size, and dust loading, are beyond the scope of this work.}

The dust-to-gas ratio mentioned several times is another important parameter {influencing the concentration mechanisms. Depending on the particle stopping time, different minimum global dust-to-gas ratios inside the system are needed to achieve unstable states \citep{yang2017}.} For the global or average ratio, we define $\epsilon$ as the ratio between the average particle mass density and the average gas density $\rho_g$:
\begin{equation}
\epsilon = \frac{\frac{N \cdot m_p}{V}}{\rho_g} = \frac{4}{3} \pi r^3 \frac{N}{V} \frac{\rho_p}{\rho_g}
\label{epsilon}
\end{equation}
Here, $\rho_p$ is still the bulk density of the individual grains (not the average solid density), $V$ is the total volume covered by particles, $N$ is the total number of particles injected, and $m_p$ is the grain mass.
The local motion of a particle will depend on other particles, as they back-react on the gas. 
Grains can be considered as "test" particles if $\epsilon \ll 1$. 
Based on a simple inertia argument, collective motion occurs at least if $\epsilon \geq$ 1.
It is one motivation of our work to study at what point exactly the density dependence of particle motion sets in while $\epsilon$ is still well below 1.

\begin{figure}[h]
\includegraphics[width=\columnwidth]{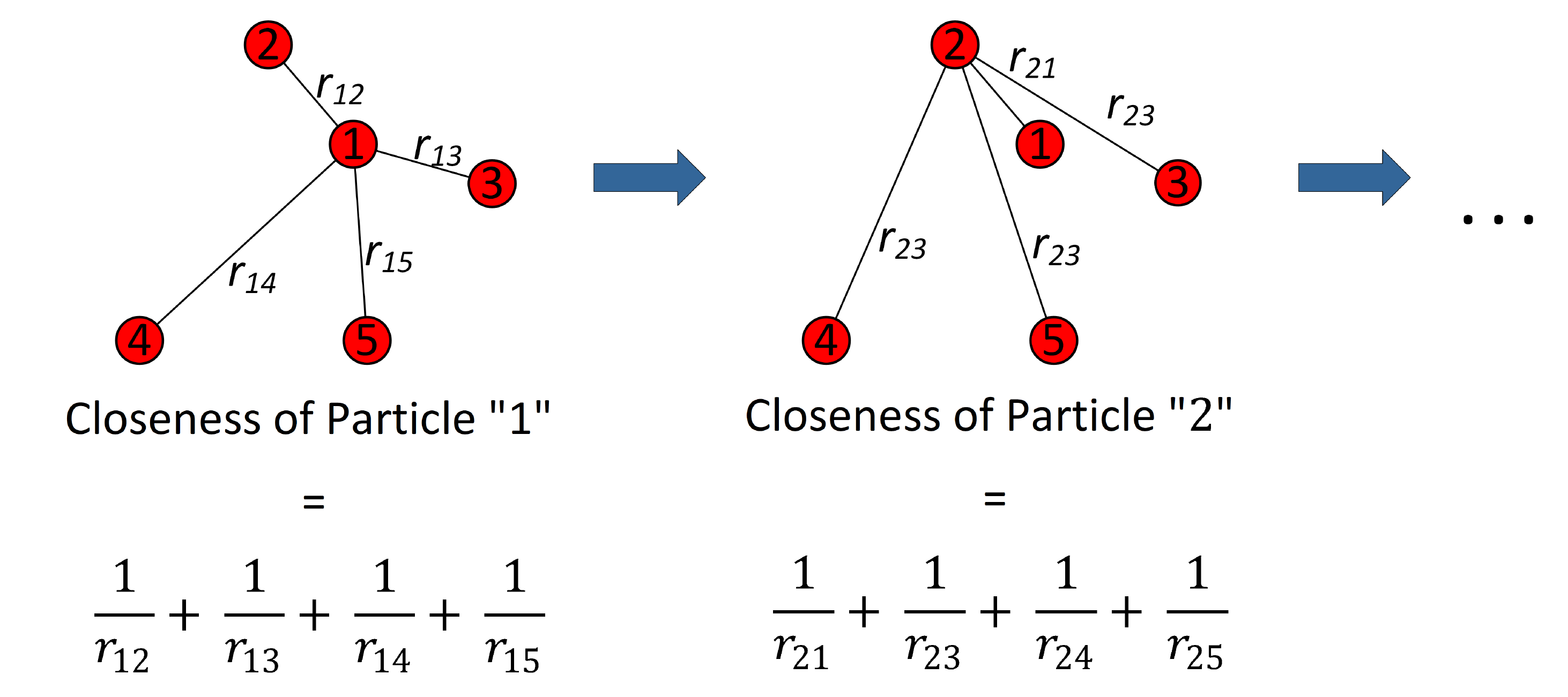}
    \caption{\label{fig.closing}On the definition of closeness {in 2D}}
\end{figure}

In particular, the local particle (or ''dust'') density will differ from the average value and change the motion of local grains. During the study underlying this paper it became clear that local density is not enough as a parameter. In hindsight, the argument would have been that for a high-density clump it would be important how large this clump is, i.e., many particles at larger distances might have influence similar to that of a few particles in a closer neighborhood. This complicates the situation, as spatial regional limits are not well defined in a fluctuating density field. We therefore introduce as a new parameter the closeness $C_i$ of a particle $i$, which accounts for both the local number density of grains and the distances between grains:
\begin{equation}
\label{eq.close}
C_i = \sum_{n=1}^N \frac{1}{r_n-r_i}
\end{equation}
{We note that because we observe particles in a 2D plane, distances between particles are also measured in a 2D space. In general, closeness is not limited to 2D.}
A graphical representation of the concept of closeness is seen in fig. \ref{fig.closing}.
Note that here, $r_i$ is the position of grain $i$, not its radius.
Closeness or valued centrality is often used in the literature on graph theory as a measure of centrality in a network \citep{marchiori2000, Dekker2005}.
The $1/r_i$ dependence in this definition is also known from the long-range hydrodynamic interactions of sedimenting spheres in Newtonian fluids under creeping flow conditions (e.g. \citet{brady1988,janosi1997,Segre1997}).
So for this first study we consider this definition of closeness as a suitable parameter to quantify the influence of a region on an individual grain's motion.

We note that closeness might not be intuitive, so we caution the reader that closeness is constructed from distances \textit{and} numbers of particles and has units of 1 $\rm mm^{-1}$. 
For our global initial dust-to-gas ratio $\epsilon =0.15$, assuming homogeneously distributed particles, the closeness would have a maximum in the center of the chamber -- we find $C = 16.7~\rm mm^{-1}$, and a minimum toward the edge of the chamber of $C =  9.3~\rm mm^{-1}$, while the mean closeness would be $C = 13.9~\rm mm^{-1}$.  The fluctuations we find in the experiments of up to $C = 40~\rm mm^{-1}$ are significant deviations from the mean value and already indicate a rather nonhomogeneous cloud.

\section{Experiment}
\subsection{Setup}

\begin{figure}[h]
\includegraphics[width=\columnwidth]{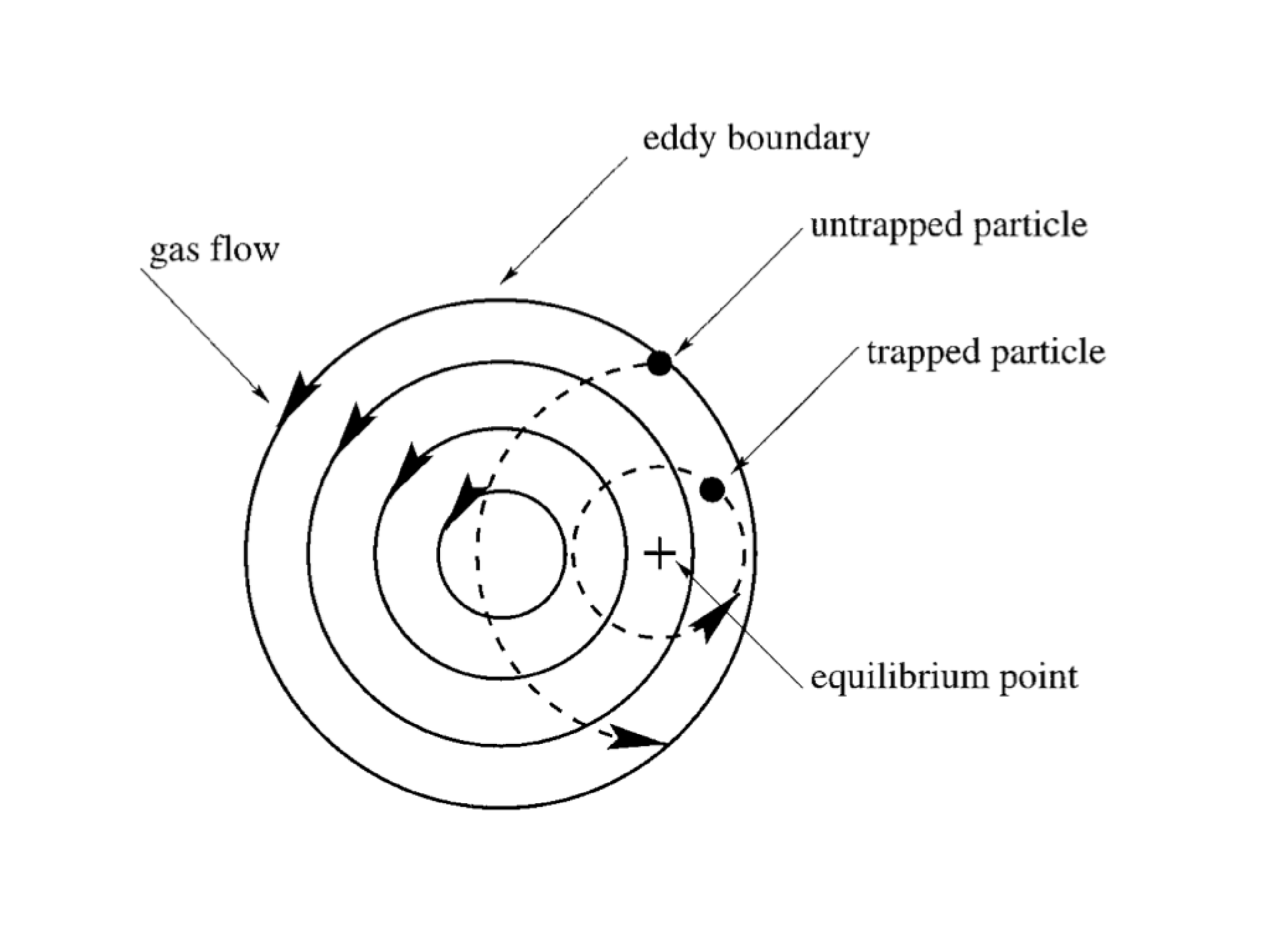}
    \caption{\label{fig.eddy} Principle of particle trapping against gravity in a convective eddy of a protoplanetary disk \citep{klahr1997}. }
\end{figure}

The basic feature of a ground-based experiment that prevents particles from sedimenting to the bottom is the trapping of these particles in circular orbits within an eddy. This idea was proposed by \citet{klahr1997} for protoplanetary disks (fig. \ref{fig.eddy}). The same idea applies in our experiment. {One difference is that the rotation of the gas is
induced not by thermal convection but by the rotation of an experiment chamber.} As the chamber walls apply friction to the gas, it
responds by rotating in a rigid rotation along with the vacuum chamber. 
%{Another difference is that in the solar nebula the vertical $z$ component (perpendicular to the disk mid-plane) of stellar gravity changes with height as $g_z = -\Omega_{Kepler}^2 z$, which leads to variations in the settling velocity. As an effect, sufficiently small particles spiral inwards towards their equilibrium point, which might be mimicked in the laboratory but is not the focus here.}

\begin{figure}[h]
\includegraphics[width=\columnwidth]{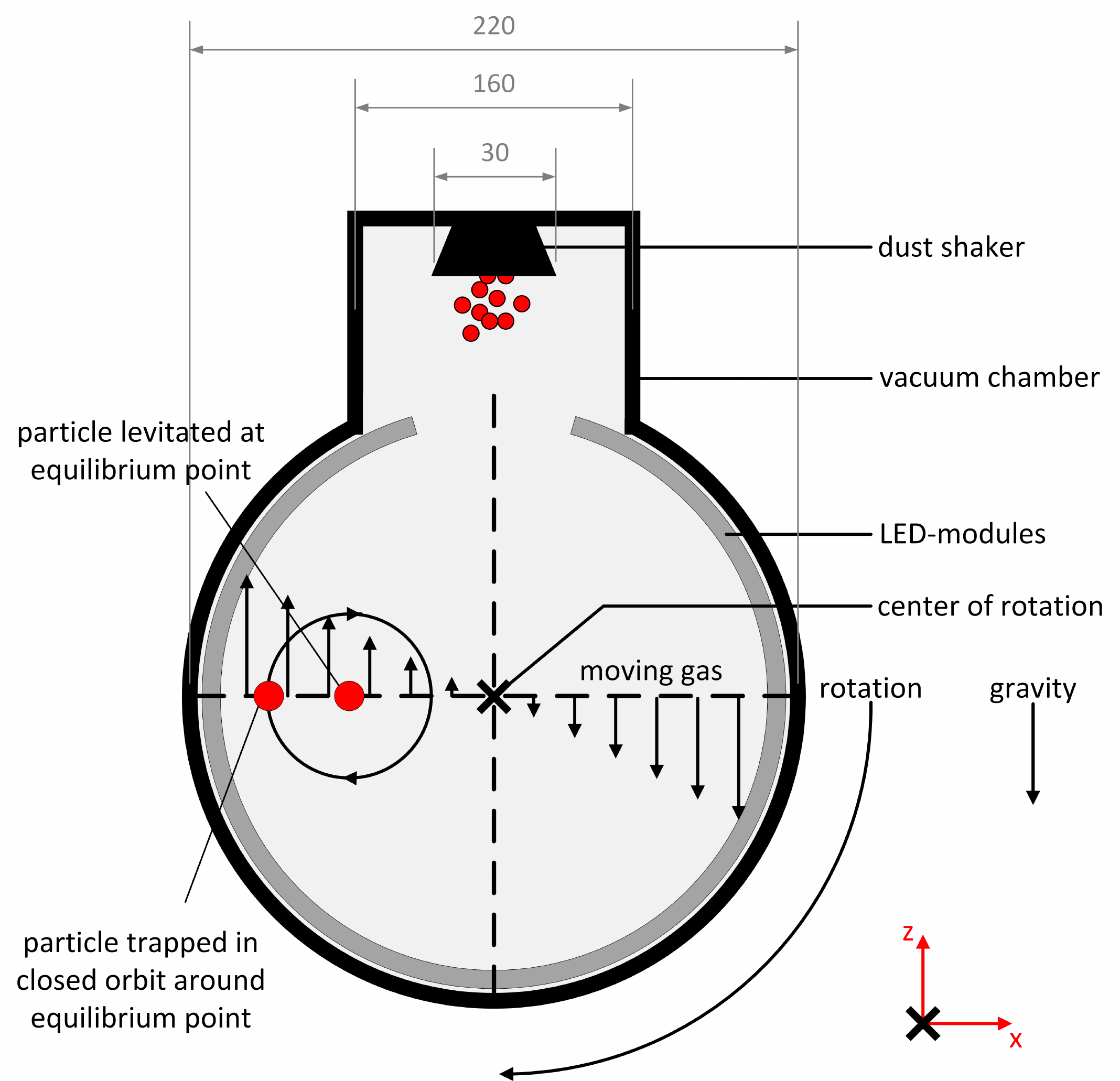}
    \caption{\label{fig.setup} Schematics of the experiment. Not shown are auxiliary parts. The experiment chamber is a vacuum chamber evacuated prior to experiments to a preset pressure. A camera observes particles from the front by scattered light. {The origin of the coordinate system is the center of rotation.}}
\end{figure}

A sketch of the experiment can be seen in fig. \ref{fig.setup}.

{The experiment chamber is about 22 cm in diameter and 25~cm in depth with the measurement plane in the middle of the chamber.} It is evacuated to a preset pressure before the experiment is started.\\
{Particles are injected into the chamber through a vibrating sieve included in
an extension of the vacuum chamber. This beam of particles has a width of 5 cm and 
a thickness of 5 mm.}
A number of electrical contacts are fed through to the rotating system. Inside the chamber a ring of LEDs generates light that is scattered from the particles, which are imaged by a non-rotating camera in the front.
{The camera observes the particles from the front at a distance of approximately 45~cm. The focal length of the objective is 35~mm. {The resolution of the camera is 1.3 megapixels ($123  \rm \mu m px^{-1}$),} frame rate is 100~fps, and the exposure time is 1800 $\rm \mu$s. The field of view is 16~x~13~cm, and the depth of field is 5~cm. }

{\subsection{Procedure}
The center of rotation of the chamber is determined by viewing the superposition of images of a grid placed inside the rotating chamber.\\
Once the experiment chamber is evacuated, the vacuum pump is disconnected.
Particles are continuously injected into the chamber while the experiment is still at rest. Due to their finite sedimentation velocity, they do not reach the bottom of the chamber immediately. Rotation is started with a preset rotation frequency $f$ and the injection mechanism is stopped. After that, recording is started.\\
Particles in regions with stable equilibrium points are trapped inside the chamber. It is necessary that the particle friction time, gas pressure, and chamber rotation frequency are well adjusted to trap particles. Real particle tracks can be seen below in fig. \ref{fig.circles}.\\
Here, the data of one of those experiments are shown. Observations are taken for 30~s. Only data after 6~s (two rotations) are considered to avoid any initial influences. Overall, about one million particle positions are tracked and analyzed.}\\

\subsection{Experimental parameters}

\begin{figure}[h]
\includegraphics[width=\columnwidth]{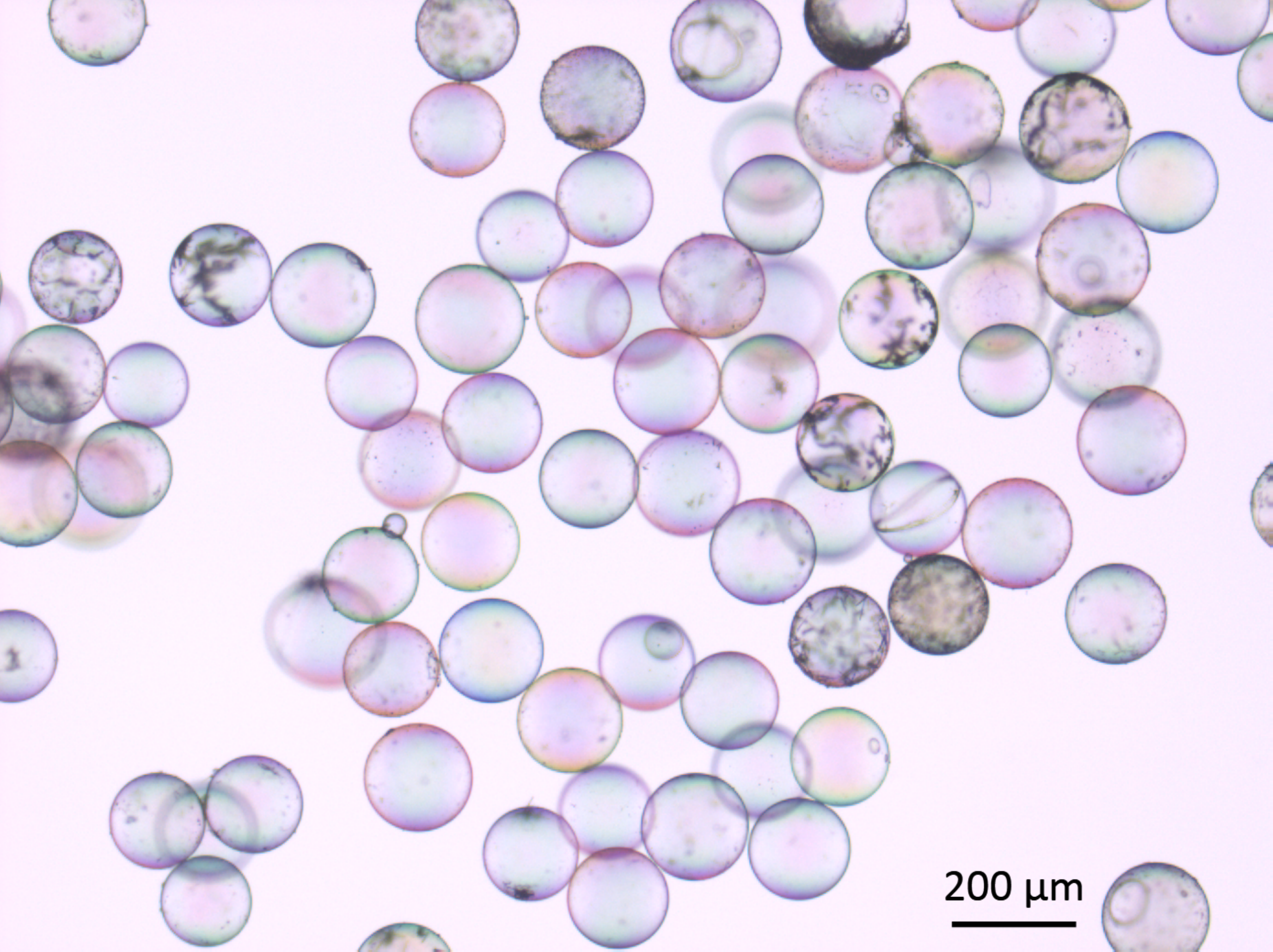}
    \caption{\label{fig.hollow}{ Microscopic image of the hollow glass spheres used. The beads are placed on a microscopic slide.}}
\end{figure}

A summary of the most important parameters of the experiment is given in table \ref{para}.
For this first study we use hollow glass spheres to get clouds of large non-sticky particles with low particle density for short gas-grain coupling times. {Here, "non-sticky" refers to the fact that the grains are of sand size (larger than about 100 $\rm \mu m$) and are not supposed to stick together easily due to surface forces upon collisions as opposed to micrometer-sized dust.} {This is of no further importance for the analysis though.} An image of the particles is shown in 
fig. \ref{fig.hollow}. 
The grains have an average diameter of 165~$\rm \mu m$. However, we will stick to terms like "dust-to-gas ratio" here for the relation between particle mass and gas mass. For the gas, we use air.

\begin{table}
\centering
\caption{Experimental parameters: Except for the grain size and all parameters deduced from it, for which the range is specified, only average values are given.}
\begin{tabular}{r|l}
Parameter&Value\\
\hline
Particle size&165 $\rm \mu m$  $\pm$ 15 $\rm \mu m$\\
Particle density&60 $\rm kg / m^3$ \\
Initial particle number& 650\\
Gas pressure&950 Pa $\pm$ 100 Pa\\
Initial dust-to-gas& 0.15$\pm$0.02 \\
Final dust-to-gas& 0.07 $\pm$0.01\\
Max. local dust-to-gas & 1.4 $\pm$0.1\\ 
Initial volume filling factor& 3 x 10$^{-5}$\\
Friction time&7 ms\\
Knudsen number&0.08\\
{Chamber} rotation frequency \textit{f} & 0.336 Hz\\
Stokes number&0.014\\
$Re_\mathrm{Setup}$ with d = 22 cm & 18.5
\label{para}
\end{tabular}
\end{table}

{
The particle size and particle density are specifications of the manufacturer. The size distribution is verified via microscopic images, but there are concerns regarding the particle density (see section 4.1).\\  
The initial and final dust-to-gas ratios are calculated using eq. \ref{epsilon} with an estimated volume of $V = 31800 \rm mm^3$, which is derived from the area of all possible stable particle trajectories and an estimated depth of 5~mm corresponding to the thickness of the injected particle beam. The local dust-to-gas ratio is calculated within a fraction of the measurement volume of $V=125\rm  mm^3$, again with a 5~mm depth.\\
Considering the depth of view of the camera, the measurement volume increases by a factor of 10, and therefore the values of $\epsilon$ would decrease by the same factor.\\
The friction time is calculated {with eq. \ref{vrel}} {considering} the terminal velocity of undisturbed sedimenting particles to prevent errors resulting from uncertainties in particle density. Therefore, the Stokes number is also not influenced by { the particle densities.} \\
{The Stokes number is calculated with the friction time and rotation frequency $St= \tau_f \cdot 2 \pi \cdot f$.}
Given that the {injected} particles move in {a} depth of 5~mm, one can give an error {caused by the 2D projection} of 1$\%$ in all quantities where the particle position is important, namely the velocities, friction time, Stokes number, Reynolds number, local $\epsilon$, and closeness.}

\subsection{Coordinate System}
%{
Although a rotating system is examined here, it is important to emphasize that the coordinate system is neither rotating nor cylindrical but Cartesian. The point of origin of the chosen coordinate system is in the center of rotation of the vacuum chamber; {the rotation has a rate $\Omega = 2 \pi f$, and the rotation axis is defined by} the $y$-axis along the horizontal plane. {The $z$-axis then is in the opposite direction of gravity (see Fig. 2 for the definition of axes). As we will see, the Ekman number of our rotating gas chamber, indicating the relative importance of rotation (Coriolis forces) in comparison to viscosity, is low, 
\begin{equation}
{\rm Ek} = \frac{\rm Ro}{\rm Re} \approx 0.025,
\end{equation}
and in the absence of any perturbations but for particle feedback leads to a rigid rotation of the
gas with
\begin{equation}
u_x = z \Omega; \,\,\,\, u_z = - x \Omega; \,\,\,\, u_y = 0.
\end{equation}
The relevant forces in the experiment are gravity $F_g = m g \vec{e}_z$ and the drag force $\vec{F_d} = - m \cdot \frac{\vec{v}-\vec{u}}{\tau_f}$, where $\tau_f$ is the friction time as detailed below. Thus individual particle motions are
approximately given by
\begin{eqnarray}
\dot v_x &=& - \frac{v_x - z \Omega}{\tau_f},\\
\dot v_y &=& - \frac{v_y}{\tau_f}, \\
\dot v_z &=& - \frac{v_z + x \Omega}{\tau_f} + g.
\end{eqnarray}
Neither centrifugal nor Coriolis forces enter this system, as it is given in non-rotating coordinates.
{In contrast to motion in protoplanetary disks, a corotation frame would not simplify the problem, as in the laboratory experiment the vector of gravity would then rotate.
Regarding the short friction time of $\tau_f=7$~ms (see below) and the rotation timescales of the experiment of about $2 \pi f = 2$~s, the solution is as follows:}

{Along the $y$-direction, the motion is a simple exponentially damped motion or $v_y = v_{y0} e^{-t/\tau_f }$. Here, $v_{y0}$ is the initial velocity in the $y$-direction. This is close to zero as particles are injected vertically. In any case, as $\tau_f$ is short, this is quickly damped and to a good approximation $v_y = 0$. The particle motion is therefore restricted to the $x$,$z$-plane.}

{Along the $x$-direction, we also have a damped motion but within a gas flow. Neglecting that $z$ varies with time due to the short friction time, the solution is $v_x = (v_{x0} + z \Omega) e^{-t/\tau_f } - z \Omega$. $v_{x0}$ would be the initial velocity in the $x$-direction, but again, the exponential part decays rapidly and the motion can be approximated by $v_x = - z \Omega$.} 

{ Last but not least, the motion in the $z$-direction is a damped freefall motion if due to the short friction time we again neglect that the $x$-position is time-dependent. The solution is
$v_z = (g \tau_f + v_{z0} + x \Omega) e^{-t/\tau_f} + x \Omega - g \tau_f$, and for fast decay the equilibrium motion is
$v_z =  x \Omega - g \tau_f$.}

{ In total the equilibrium solutions are
\begin{eqnarray}
v_x &=& -z \Omega,\\
v_y &=& 0, \\
v_z &=& x \Omega - g \tau_f .
\end{eqnarray}
}

{which is rotation at rate $\Omega$ shifted by $dx = \frac{g \tau }{\Omega}$ from the rotation center of the gas.
}
As a result, single unperturbed particles either levitate at an equilibrium point or are trapped in closed orbits around  this point.}

{{While the problem is not treated in a rotating frame, the approximations behind these calculations can still be quantified by considering centrifugal forces. At the equilibrium point they are zero. The centrifugal forces are otherwise still orders of magnitude smaller than the gravitational acceleration $\Omega^2 r << g_z$ (see Section 2). Also, the radial outward drift due to centrifugal forces $v_r = \tau \Omega^2 r$ would bring a Coriolis force, but as a second-order correction with $F_{\rm Cor} = 2 \tau \Omega^3 r$, this is even smaller than the centrifugal term. Using the numbers of the experiment for $\Omega$ and $\tau$ and for a radius of $0.1$ m, we get a ratio of gravity to centrifugal force to Coriolis force of about 10 / 0.5 / 0.01.
The calculations above are therefore exact to about 5\% in the context of the experiments.}}

{\subsection{Analysis}}
{Image editing and particle analysis are done with ImageJ \citep{fiji2012}. For the analysis, individual grain positions and tracks in a Cartesian coordinate system are generated by the TrackMate plug-in \citep{trackmate}. 
{Particles are not resolved in 3D}. In case of two overlapping particles, one track is cut but continued if the particle appears again. {In high particle loading, particles might be missed. However, given that even in this case the cloud is far from being optically thick, these are rare events which we consider to be of no significance here.} A wrong assignment can be prevented by customizing the parameters of the particle-linking processes. For our analysis, particle track length is not important due to the large amount of data averaging out details, with most cases allowing determination with low errors on the percent level. The same is true for particle position and closeness (see below), which is not sensitive to and does not require a detailed search for the best individual particle position algorithm.}
Velocities are determined from short parts of these tracks.

{With the help of the TrackMate feature, particle positions are assigned from frame to frame. We assume that the particle velocity between frames is constant, and calculate the particle velocity for every frame and every particle. With the known center of rotation and position of the particle at the given frame, the gas velocity at that point is calculated by assuming rigid rotation. From the gas velocity at the current particle position and particle velocity, the relative velocity, which we will refer to as sedimentation velocity, is determined.
The rigid rotation is validated by comparing the sedimentation velocity of particles at low particle loading (to prevent particles from influencing one another) for different rotation frequencies and therefore different equilibrium points. These shift in a linear way, so that a linear increase from the center to the wall of the chamber is appropriate.}

{Closeness is calculated for every particle using equation \ref{eq.close} with all particle positions at the given frame}.

\begin{figure}[h]
\includegraphics[width=\columnwidth]{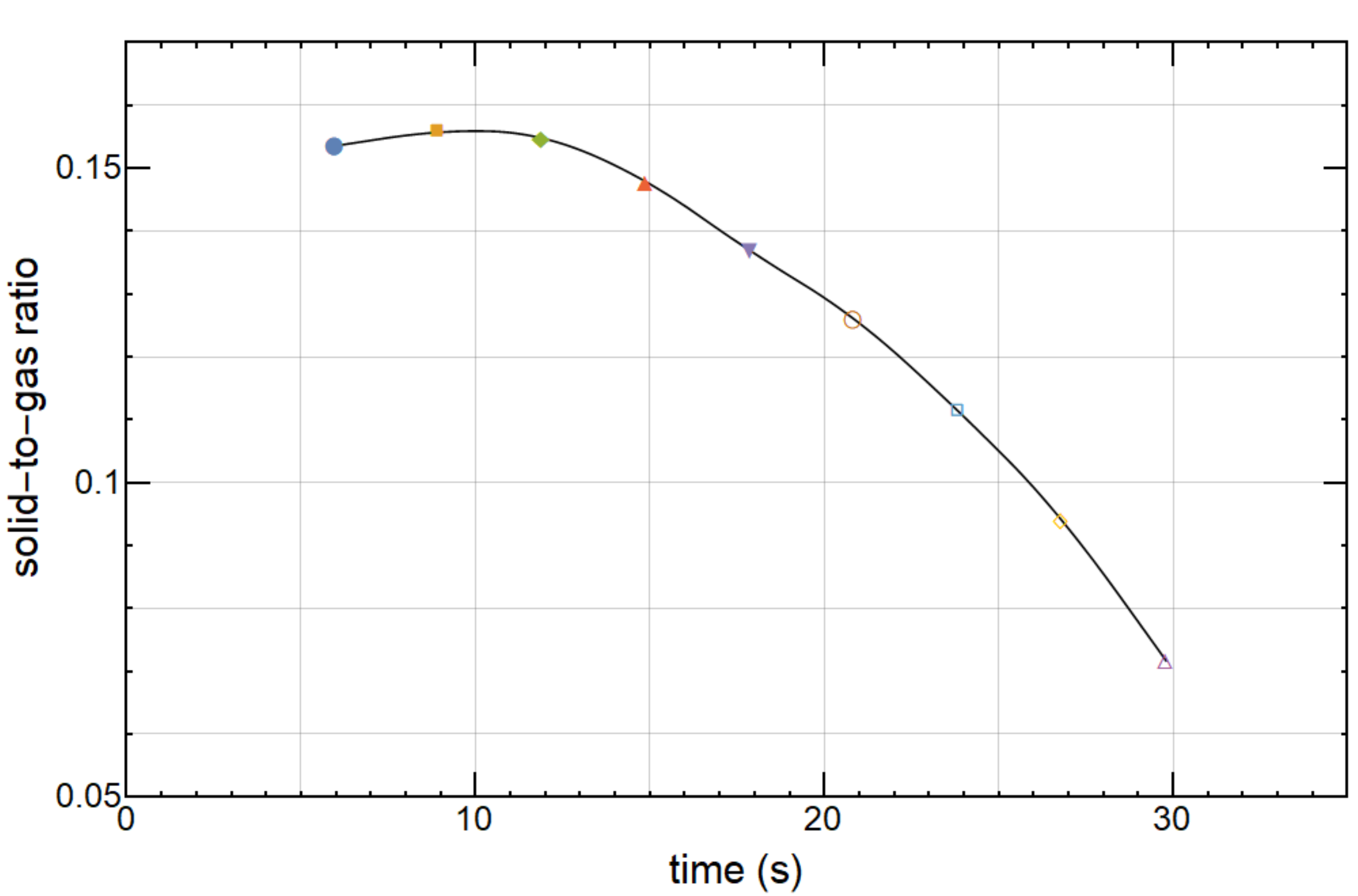}
    \caption{\label{fig.dtog}Global dust-to-gas ratio over time. Dust density is calculated with respect to the depth of particle injection. {The color and shape of the data points correspond to those of the points in fig. \ref{fig.vvonc} }}
\end{figure}

For the average dust-to-gas ratio $\epsilon$, we take the particle mass over the gas mass within the volume initially filled with particles. 
{Mainly due to increased sedimentation velocity in high local particle loadings, particles increase their orbit and collide} with the chamber walls. {Following from this, particles are lost in the course of the experiment, and} the particle density decreases with time. {Therefore, all data points are {grouped in} single full rotations of the chamber.
{The total number of evaluated data points is $\approx\!\!875,000$; each full-rotation point represents between $\approx\!\!54,000$ and $\approx\!\!118,000$  data points.}
The shape and color of the data points match the respective rotation.}
{The decreasing solid-to-gas ratio due to lost particles} is shown in fig. \ref{fig.dtog}.
In agreement with the estimates above, significant changes occur in 10 orbits, or 30 s. Details for clumps of different closeness are given below.

It has to be noted that these dust-to-gas ratios are calculated based on the average grain size, the observed number of grains, and the 5 mm thickness of the injected particle beam.
If the beam were to disperse along the cylinder axis over time, then the low densities at later times would be lower.

\section{Grain Motion}

\subsection{Grain Motion at Low Particle Loading}

\begin{figure}[h]
\includegraphics[width=\columnwidth]{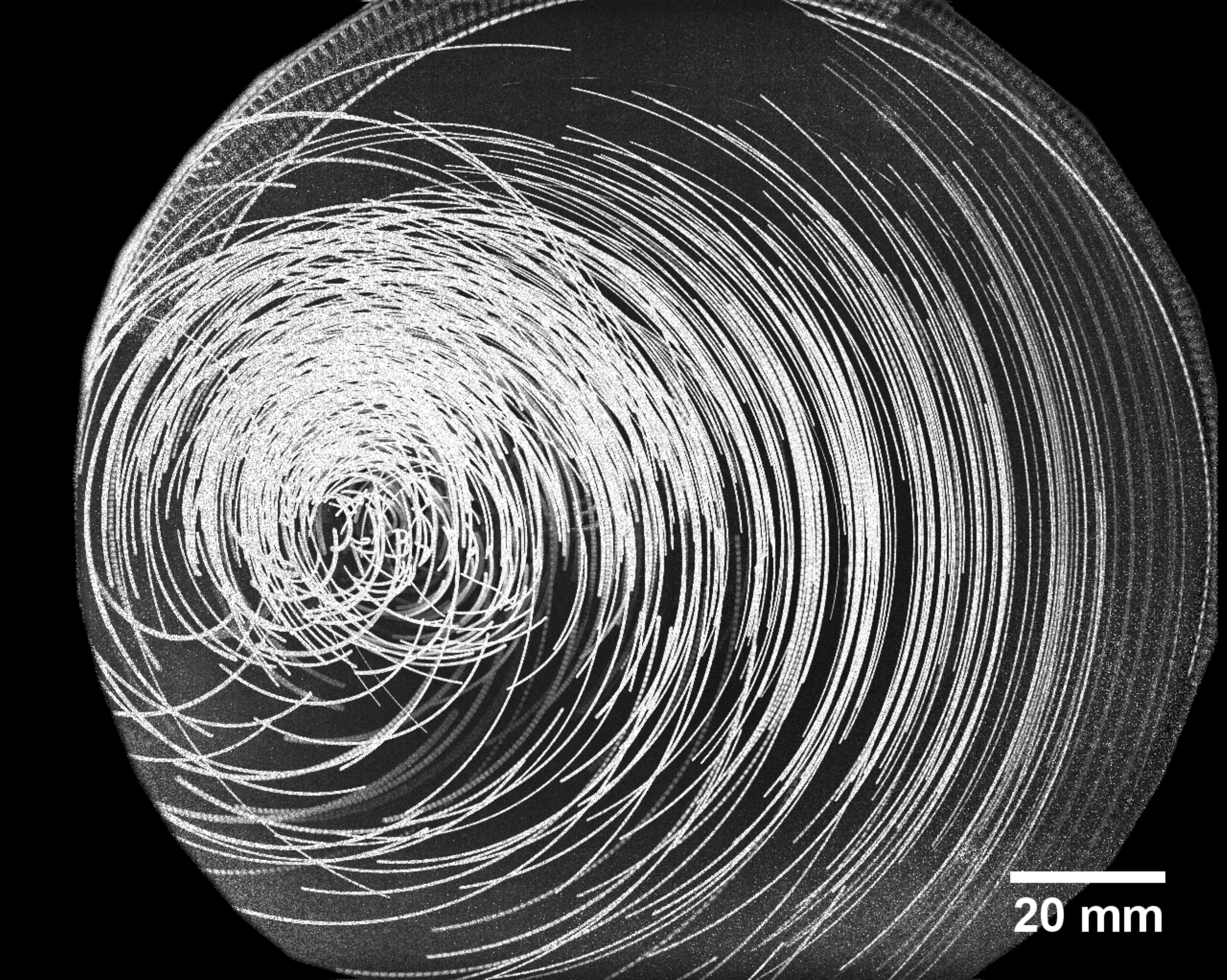}
    \caption{\label{fig.circles}{ Particle tracks as superposition for one-fourth of a rotation. The chamber rotates clockwise. Due to variations in the sedimentation speeds studied, there is a range of equilibrium positions. Therefore, the circular tracks have different centers.}}
\end{figure}

Especially at low dust-to-gas ratios, the absolute gas motion coincides with the rigid rotation around the center with the set rotation frequency.
For all particles we calculate a relative velocity by assuming the gas to be in such a circular motion and subtracting this absolute motion. {Centrifugal parts are neglected.}
Grains in steady-state sedimentation should move in circular orbits in the laboratory 
reference frame, with $v_{rel} = \tau_f g$ relative to the gas. Indeed this is the case. 
Fig. \ref{fig.circles} shows the motion of individual particles.
Due to the limited variation in particle size the point of stability, which
is the center of these circles (not the center of the chamber), is similar for all particles.
The calculated sedimentation velocity {(eq. \ref{tauS} and \ref{vrel})} for an average particle is 50 mm/s, but this bears uncertainties, e.g. in {volumetric mass density (see Table \ref{para})}, which is not well constrained. We therefore determine the sedimentation velocity of a single grain by dropping individual grains in a gas of the same pressure used here and find a sedimentation velocity of 69 mm/s, roughly consistent with the calculated value. This value for a single grain fits well the observed sedimentation velocities of the rotating grains at low particle loading, with an average of 
68~mm/s (see fig. \ref{fig.vvonc} below). The closeness in this case of the lowest $\epsilon$ reaches up to 20 $\rm ^{-1}$ but has no influence on the average sedimentation velocity.\\
{Comparing the measured and calculated sedimentation velocities, a bulk density of approximately 75 $\rm kg / m^3$ would fit to the real terminal velocity rather than the manufacturer specification of 60 $\rm kg / m^3$. In this regard, the dust-to-gas ratios given in the paper are underestimated by a factor of roughly 25$\%$.}

\subsection{Grain Motion at High Particle Loading}

\begin{figure}[h]
\includegraphics[width=\columnwidth]{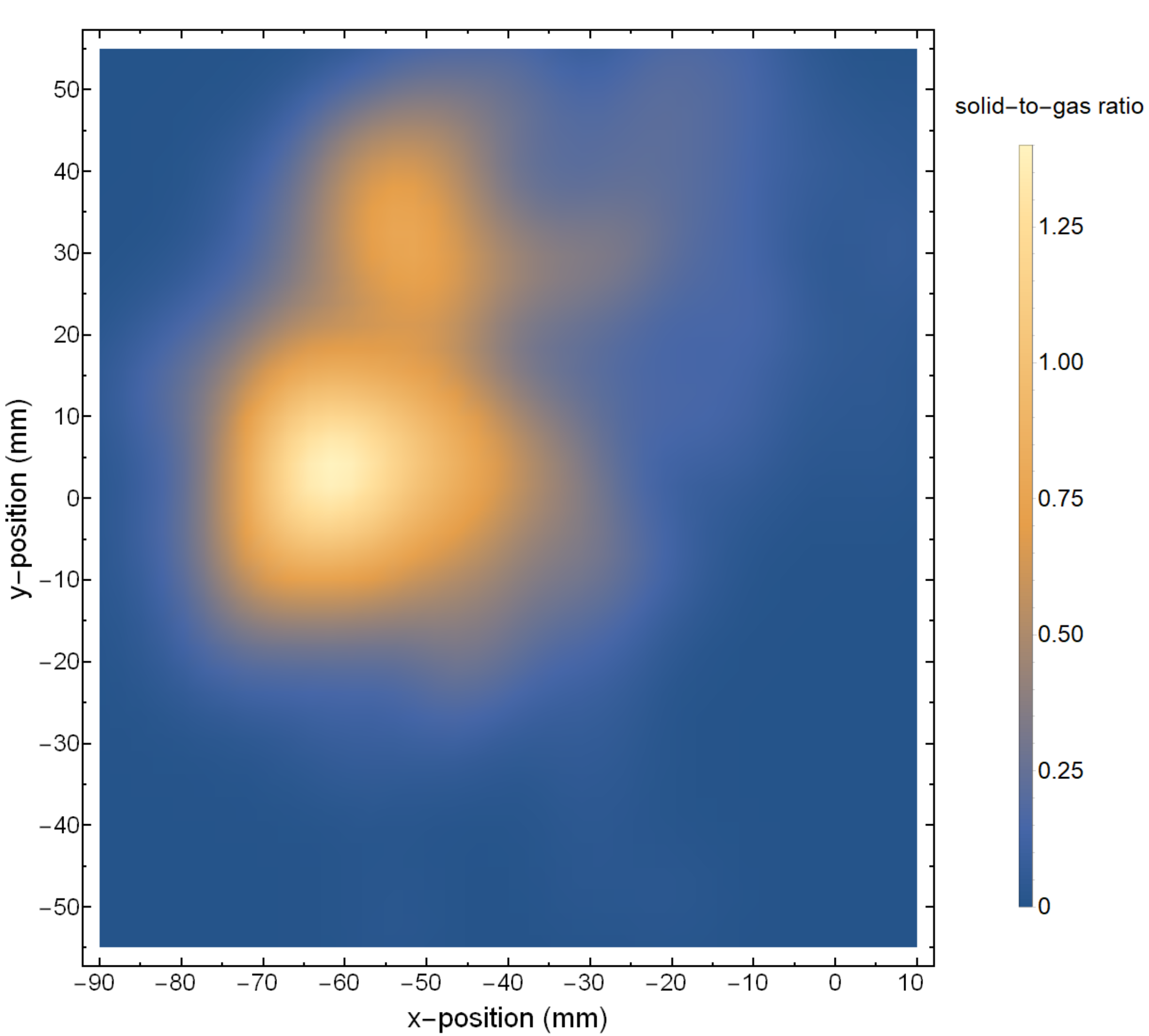}
    \caption{\label{fig.dtog2}Snapshot of local dust-to-gas ratio at early times (dense state) based on 2D projections. 
}
\end{figure}

\begin{figure}[h]
\includegraphics[width=\columnwidth]{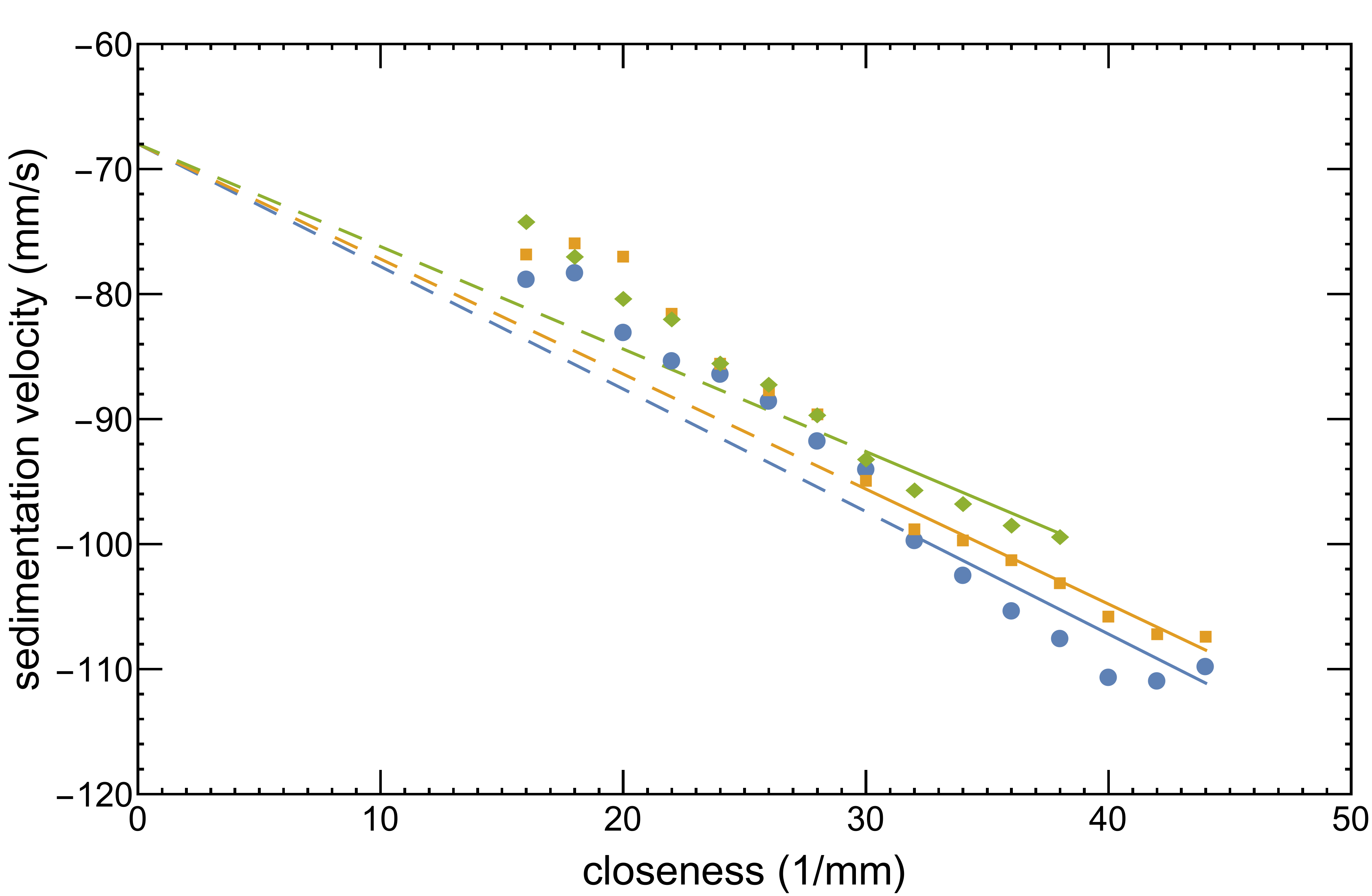}
\includegraphics[width=\columnwidth]{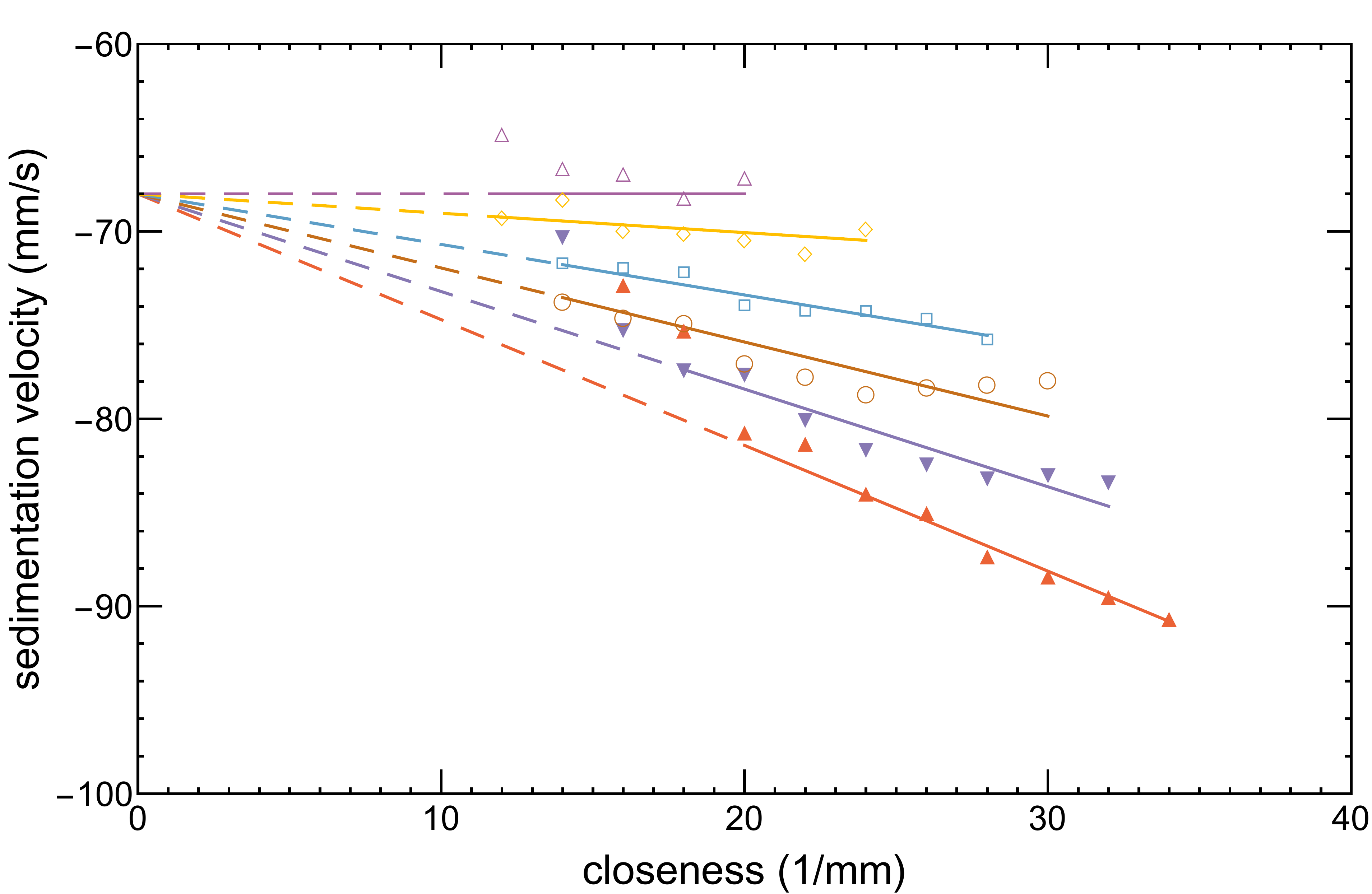}
    \caption{\label{fig.vvonc} Sedimentation velocity over closeness for individual rotations of the experiment chamber. The evolution with time goes from initially dense clouds { to less dense clouds. The upper graph shows rounds 2 - 4 (lowest curve, second round; topmost curve, fourth round); the lower graph shows rounds 4 - 10 (lowest curve, fourth round; topmost curve, tenth round)}. The data are the average values for at least 1000 particle positions {with equidistant spacing of the binned values in closeness space}. The mean number of values contained in one binned data point is 7500. {The standard errors of the mean values vary between 0.1\% and 0.6\%.}
	{For rounds 2 - 4, the lines are the linear fits for the high-closeness data points with a vertical intercept of 68 mm/s. For rounds 5 - 10, the lines are the linear fits with a vertical intercept of 68 mm/s. The non-dashed lines indicate the range of data points used for the fitting. The top line (round 10) in the lower figure is a straight line as the sedimentation velocity cannot be lower than the sedimentation velocity of a single particle. There are a few outliers at lower closeness for each round. These are due to a fraction of small grains present, which sediment slower.}
    }
\end{figure}

After injection, some particles always get lost, decreasing the overall number of grains.
While the average dust-to-gas ratio $\epsilon$ only varies by about a factor of two between the beginning of the experiment (high ratio) and the end (low ratio), 
the variation in particle motion is much stronger.
Fig. \ref{fig.dtog2} shows a snapshot of the variations of local dust-to-gas ratios at early times, with values up to $\epsilon_{\rm local} \geq 1.4$.
Particles in regions with high closeness now sediment 
much 
faster than particles in less
close regions, as seen in fig. \ref{fig.vvonc} (lower data). This figure shows the sedimentation velocities and closenesses averaged over one full rotation each.  This time sequence correlates to a change in the global $\epsilon$ (see, e.g. fig. \ref{fig.dtog}, which uses the same symbols). 
The solid-to-gas ratio decreases from about $\epsilon = 0.15$ at round 2 or 6~s 
(lower curve in fig. \ref{fig.vvonc}) to $\epsilon = 0.07$ at round 10 or 30~s (upper curve), as seen in fig.~\ref{fig.dtog}.
{The outliers with lower sedimentation velocity at low closeness are due to small grains. They can reside in regions not accessible for larger grains.}

As %described above and
seen in round 10 (fig. \ref{fig.vvonc}), at later, less dense times,
all particles sediment with the same speed, also at a closeness of 20 $\rm mm^{-1}$ .
In contrast, in the high loading case, the speed of particles with low closeness also 
increases with closeness. Obviously the system becomes sensitive to closeness and closeness variation only 
in the denser case.

\begin{figure}[h]
\includegraphics[width=\columnwidth]{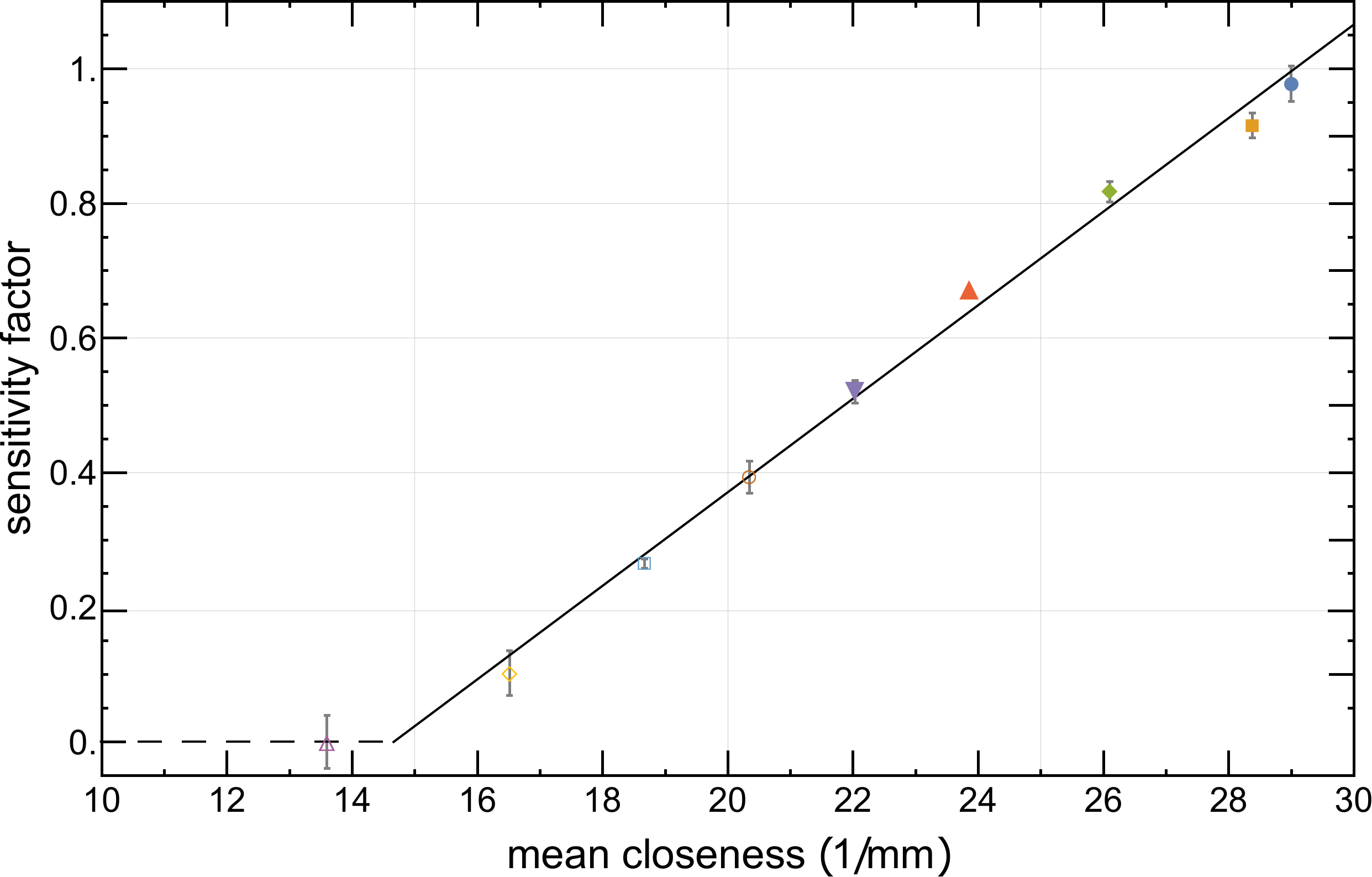}\par
\includegraphics[width=\columnwidth]{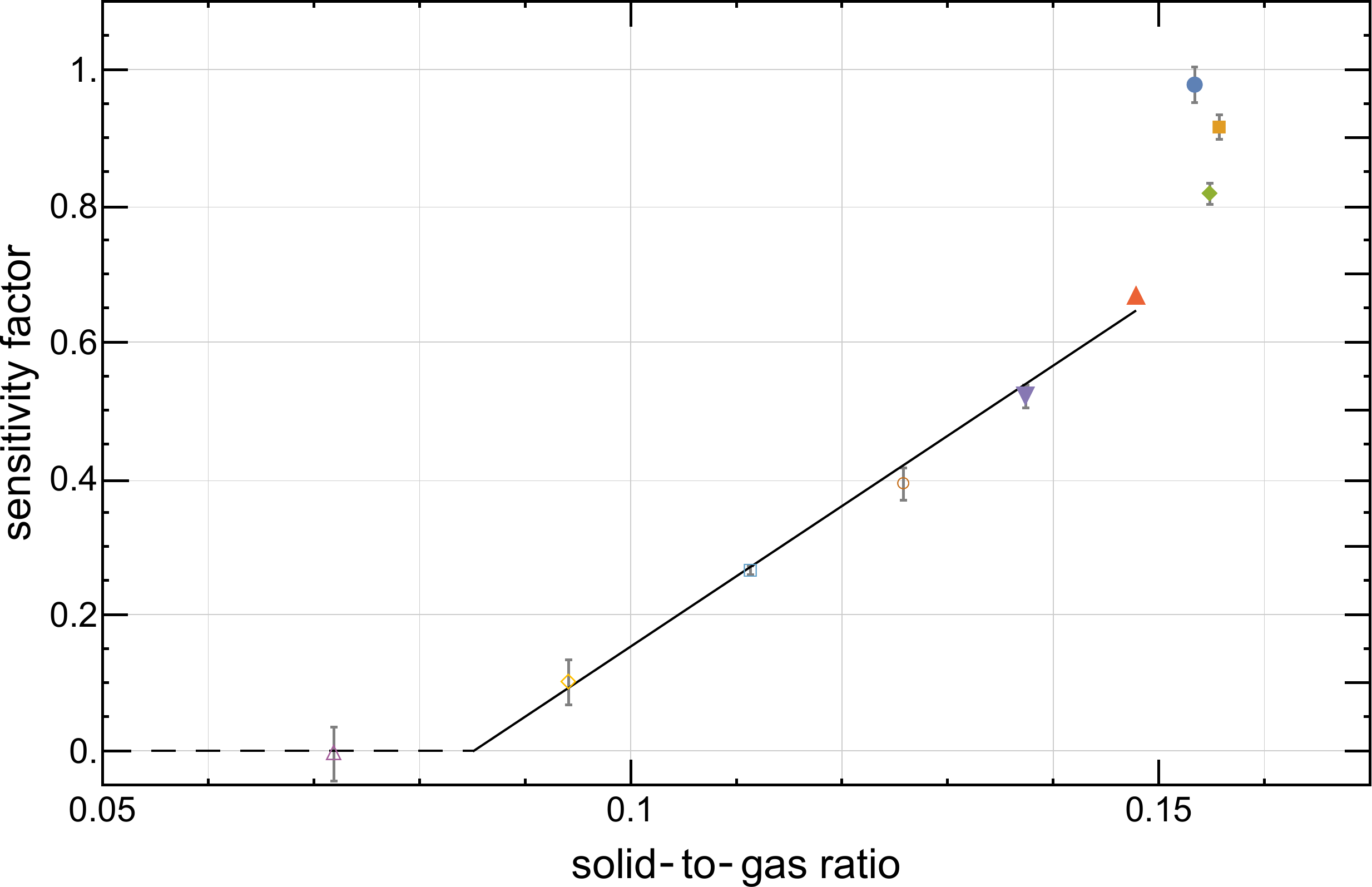}
    \caption{\label{fig.sensitive1} Sensitivity factor (in $\rm mm^2 s^{-1}$) dependence on average closeness (top) and dust-to-gas ratio (bottom); {the lines are linear fits with slopes (top) $\alpha = 0.069 \pm 0.002\,  \rm mm^3/s$ and (bottom) $\alpha = 10.3 \pm 0.5\, \rm mm^2/s$. {The offset of the linear fits is about  $14 \pm 1\rm mm^{-1}$ (top) and $0.08 \pm 0.01$ (bottom).} {The error bars show the error of the sensitivity factors of the linear fits in fig. \ref{fig.vvonc}.}
    {We draw attention to the fact that the sensitivity factor for high closeness($> 25\rm mm^{-1}$) and high solid-to-gas-ratios ($>0.15$) is only valid for the ranges indicated in Figure \ref{fig.vvonc}.}}}
\end{figure}

As a first approximation, the dependence of the sedimentation speed on closeness can be described as linear or
\begin{equation}
\label{eq.vcloseness}
v = v_0 - F_s \cdot C_i
\end{equation}
{This slope $F_s$ corresponds to the lines in fig. \ref{fig.vvonc}}.
Note that $v$ and $v_0$ are negative in our notation of sedimentation.
{The sensitivity factor $F_s$ is not constant over time.} It increases with the average closeness or dust-to-gas ratio of the system, as shown in fig. \ref{fig.sensitive1}.
{The sensitivity factor cannot be lower than 0. This is due to the fact that particles in a dilute configuration in our system can never sediment at a rate slower than their undisturbed sedimentation velocity $v_0$. The turnover point is determined by the crossing of the x-axis of the linear fit (black line) in fig. \ref{fig.sensitive1}.}

{Therefore, in total the sedimentation velocity can be expressed as
\begin{equation}
v = v_0 + \alpha \cdot (\epsilon-\epsilon_{\rm{crit}}) \cdot C   
\end{equation}
if $\epsilon$ is higher than the critical $\epsilon_{\rm crit}$.} {The value of $\alpha$ is about 10.3~$\rm{mm^2 s^{-1}}$ for the given experimental conditions.}

If the total cloud is sensitive (high $\epsilon$) and if a region of high closeness forms or approaches a particle, this particle can join this region of high closeness and sediment faster. This is in contrast to thinner clouds, where all particles essentially move independently. 

\begin{figure}[h]
\includegraphics[width=\columnwidth]{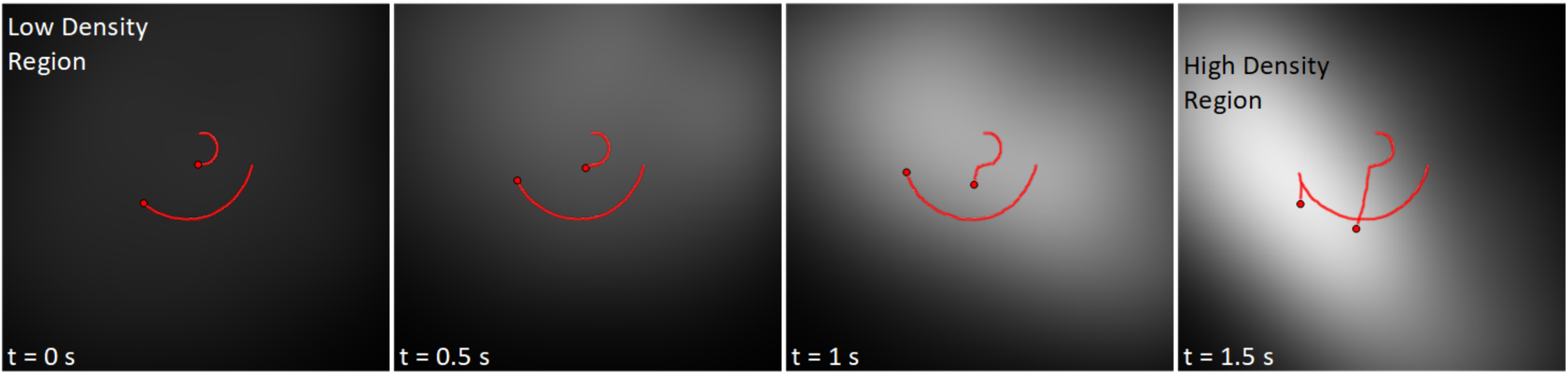}
\includegraphics[width=\columnwidth]{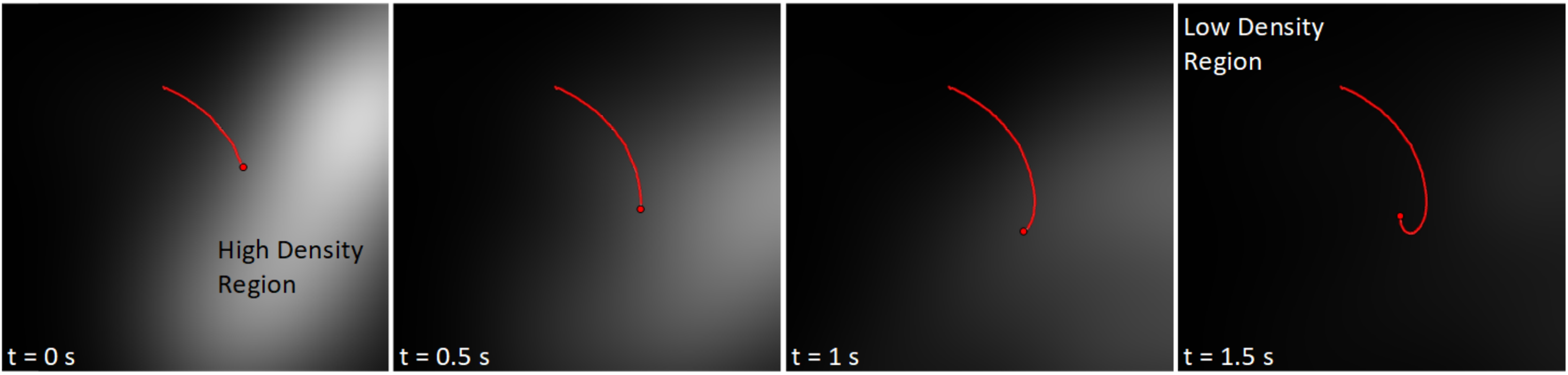}
    \caption{\label{fig.entrain1} Top: particles entering a region of high closeness getting entrained; bottom: particle leaving a high-closeness region staying behind. {A dark background color corresponds to a low particle density region and a light background color to a high particle density region. The trajectories are colored red for better recognition. The images originate from real measurements and show a 20 x 20~mm image section.}}
\end{figure}

This does not imply an instability with a steadily growing particle number as grains can also drop out of high-closeness regions. 
The trajectories of such particles are shown in fig. \ref{fig.entrain1}.

% Fig. \ref{fig.track1} shows the evolution of closeness attributed to a single particle over time and the corresponding sedimentation speed. Closeness is the main parameter influencing single particle trajectories, though other not yet examined influences lead to small deviations in sedimentation velocity from an average behavior.  This might also be related to the unknown 3d closeness. As such, non-monotonous variations might also result in loops as seen for the particle in fig. \ref{fig.track1} .

\begin{figure}[h]
\includegraphics[width=\columnwidth]{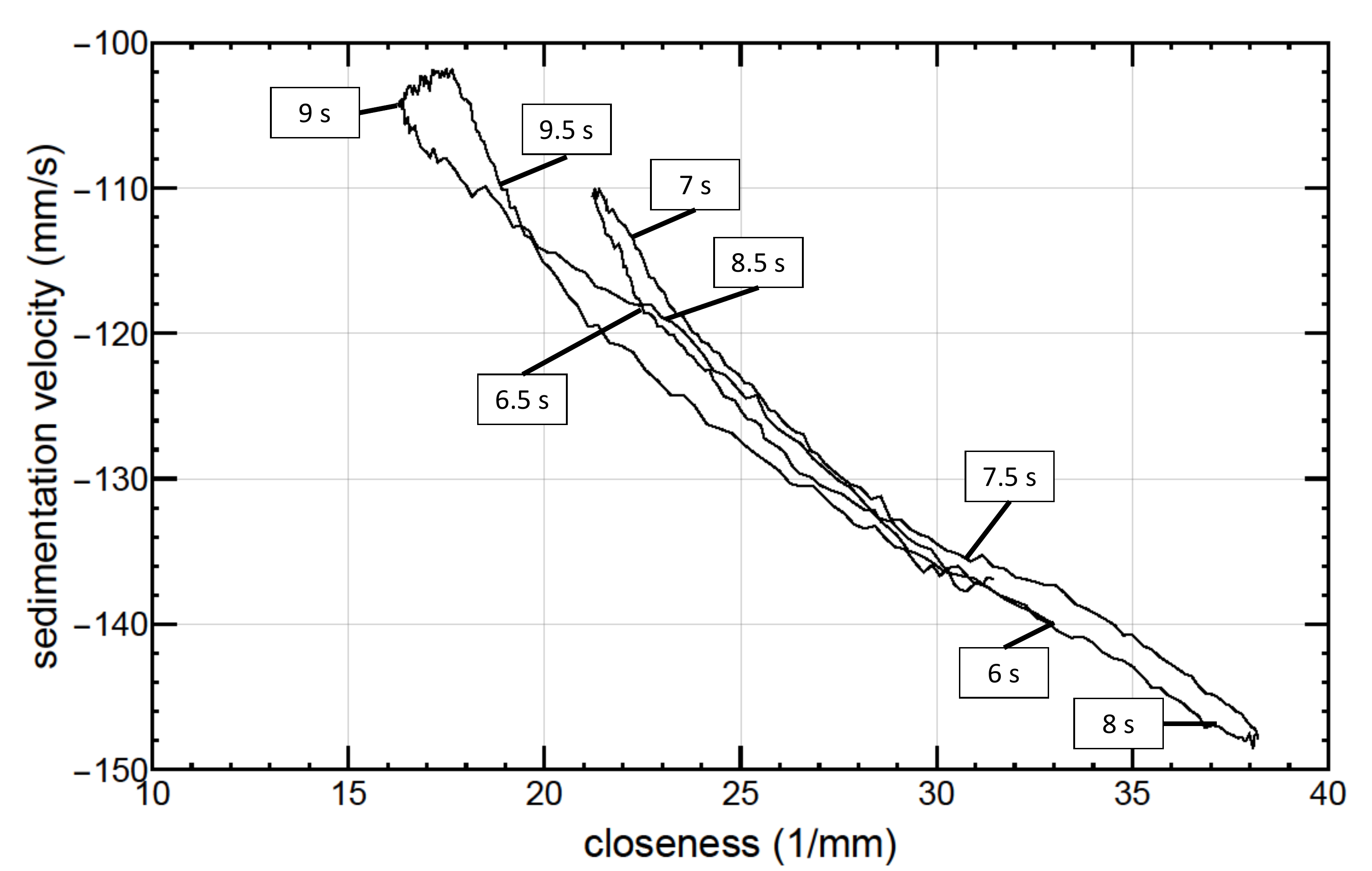}
    \caption{\label{fig.track2} {Example of an individual particle in a dense cloud in a closeness-velocity diagram over about one round. The particle enters and leaves dense areas multiple times. The times refer to the absolute time of the experiment (see fig. \ref{fig.dtog})}}
\end{figure}

Fig. \ref{fig.track2} shows the evolution of the closeness of {a} single particle over about 4~s (slightly longer than one round). Compared to the mean sedimentation velocity, the particle has a higher velocity. This is caused by the biased choice of the particle. Larger particles have higher visibility and therefore can be tracked for a longer period. 
The particle starts at high closeness (6~s), enters a less close particle configuration (7~s), re-enters a high-closeness area (8~s), leaves it again (9~s), and increases in closeness until the end of the track.
{Closeness is the main parameter influencing single-particle trajectories, though other, not yet examined influences lead to small deviations in sedimentation velocity from the average behavior.  This might also be related to the unknown 3D closeness. As such, non-monotonic variations might also result in small loops.} It is obvious that the {closeness-velocity relation} is not caused by the experimental setup or procedure.
{A main finding is that the sedimentation velocity of individual particles is linearly dependent on closeness, which is also the main motivation for our model shown in eq. \ref{eq.vcloseness}.}

Particles can join and {re-enter} dense particle configurations multiple times on timescales shorter than the period of one rotation of the experiment (fig. \ref{fig.track2}). 

% \begin{figure}[h]
% \includegraphics[width=\columnwidth]{track.pdf}
%     \caption{\label{fig.track1} Example of an individual particle in a dense cloud in a closeness-velocity diagram.{ The particle starts at high closeness at about 7~s and ends at lower closeness at 12~s.}}
% \end{figure}

\begin{figure}[h]
\includegraphics[width=\columnwidth]{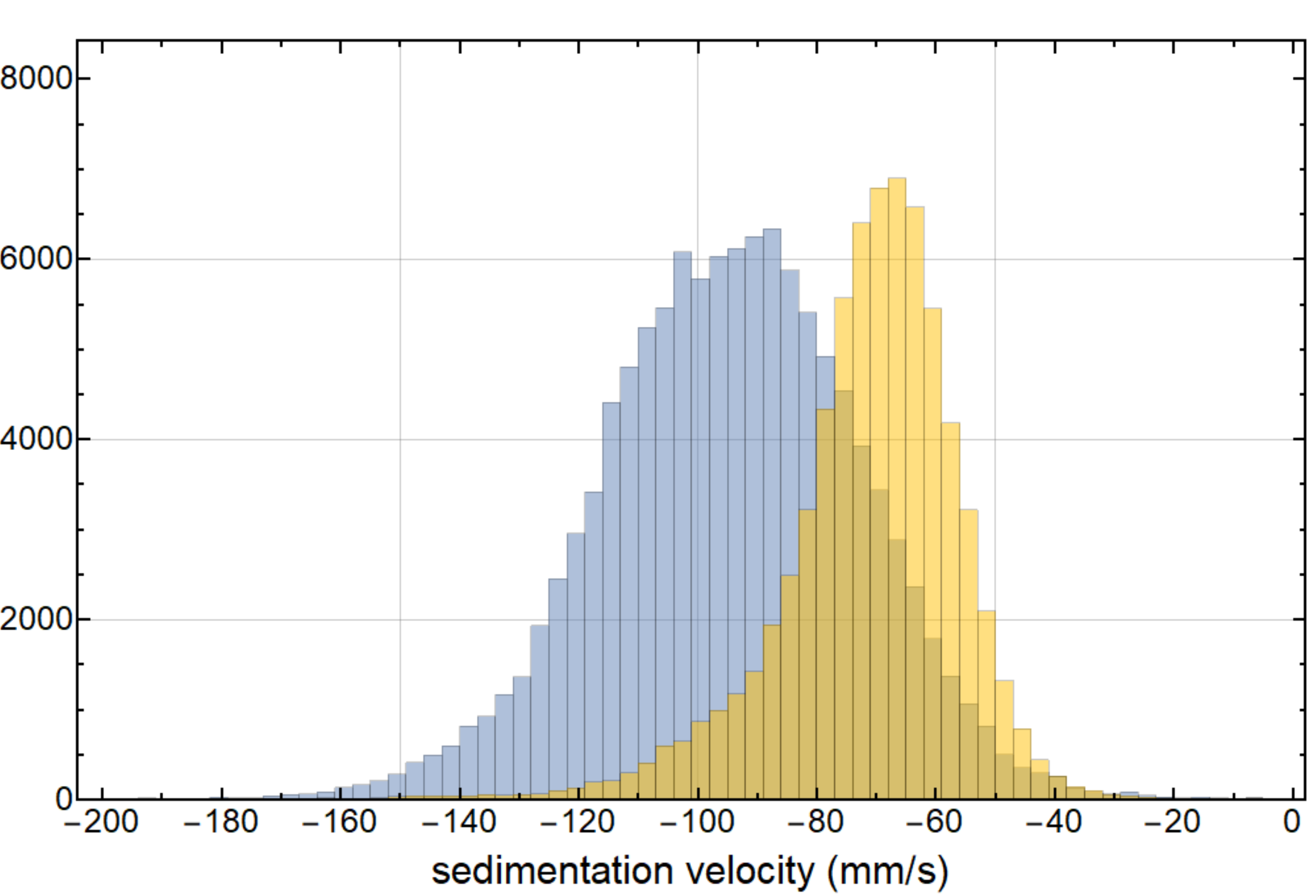}
    \caption{\label{fig.speed1} Histogram of the sedimentation speed for low particle loading (right distribution) and high particle loading (left distribution). {The $y$-axis represents the count of data points of the particles with the corresponding sedimentation velocity.}}
\end{figure}

At late times all particles on average sediment like individual grains with the calculated sedimentation velocity. However, the speeds still vary due to variations in particle size.  Typical velocity distributions for the low-density case are seen in fig. \ref{fig.speed1} (right distribution).
At earlier times (time of 6~s, round 2) variations due to an increased sensitivity are added, and the variations are much larger as seen in fig. \ref{fig.speed1} (left distribution). Note that the lower values due to the negative values are the higher absolute sedimentation velocities. 

\begin{figure}[h]
\includegraphics[width=\columnwidth]{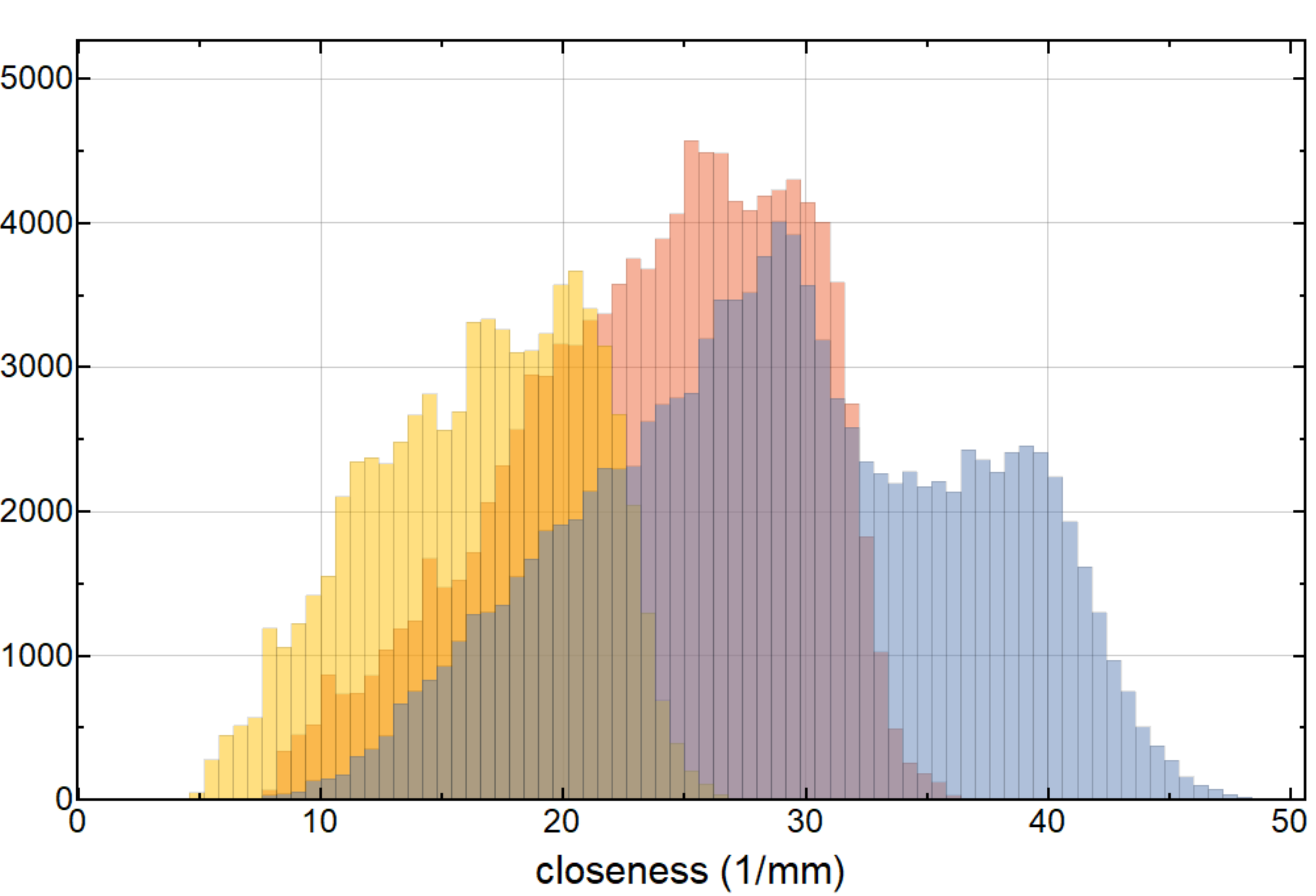}
    \caption{\label{fig.speed2} {Histogram on the evolution} of the closeness distribution for three different times from right ({early/high particle loading}) to left  ({late/low particle loading}). {The $y$-axis represents the count of data points of the particles with the corresponding closeness.}}
\end{figure}

Fig. \ref{fig.speed2} shows the closeness distribution changing from late to early times.
As can be seen, the initial closeness distribution has a wing at the high-value side, which it loses first. In general, particle regions with high closeness are preferentially lost, leading to an inclined distribution with a steeper dropoff to higher values. 
This is due to the fact that particles sedimenting faster have a smaller stable region to rotate within in the experiment and are more easily lost to collisions with the wall.

\begin{figure}[h]
\includegraphics[width=\columnwidth]{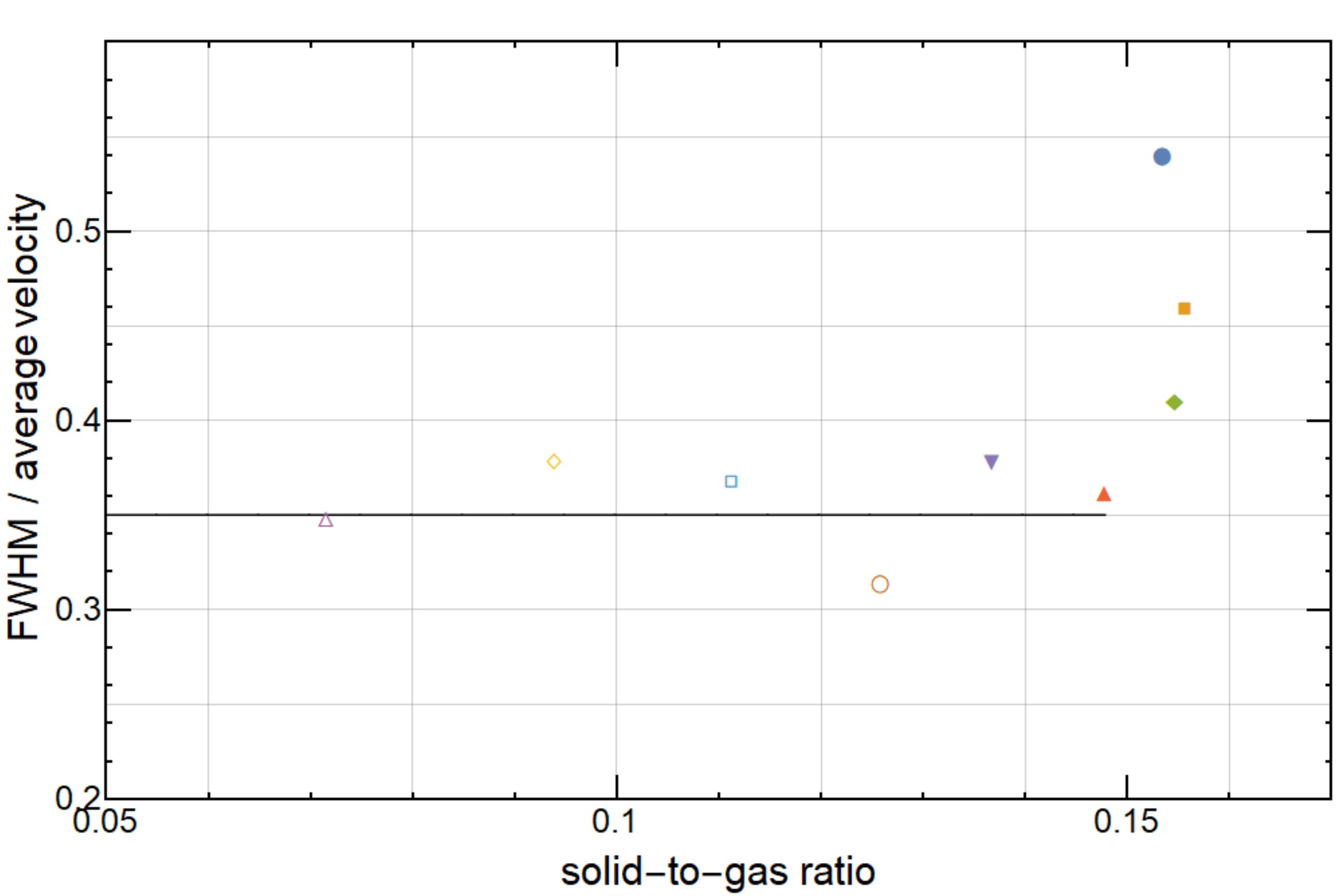}
\includegraphics[width=\columnwidth]{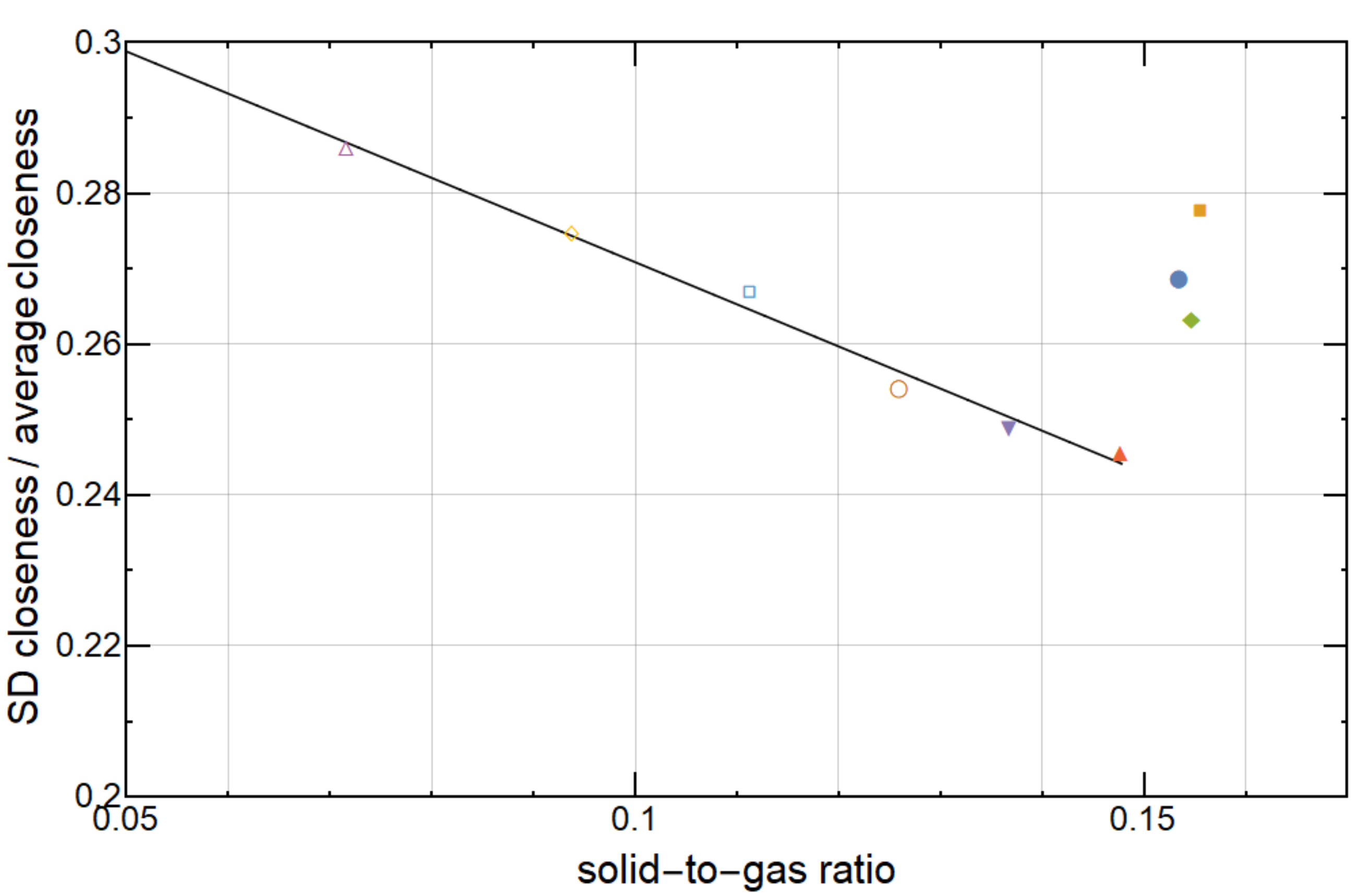}
    \caption{\label{fig.fluctuate} FWHM of the relative velocity {distribution divided by the} average velocity, and standard deviation (SD) of the  closeness {distribution divided by the average closeness over the dust-to-gas ratios. The FWHM is used because of the almost Gaussian shape; the SD is used because the closeness distribution is not Gaussian-shaped (see Fig. \ref{fig.speed1} and \ref{fig.speed2}).}}
\end{figure}

{Some of this is summarized in fig. \ref{fig.fluctuate}.}
It shows that the relative width in velocity essentially stays constant, but the initial fluctuations at high dust-to-gas ratios are much broader. {The relative width at later times (lower dust-to-gas ratios) is only somewhat larger than the size distribution of about 20\% full width. Otherwise, this is similar to the findings in other sedimentation experiments. For example, \citet{guazzelli2011} found 
an increase of the relative width with the filling factor (proportional to the dust-to-gas ratio). However, the dependence they found is very weak at the low filling factors relevant for our experiment. 

In any case, our data imply that in a region of higher closeness, where the average behavior changes and the sedimentation velocities also increase, the variation in speed increases in the same manner. {Therefore, while the absolute sedimentation velocity increases in a dense region, the sedimentation speed is still linear to particle size. In other words, if a grain of average size is a certain factor faster, also smaller or larger grains in the same dense region are faster by the same factor.} This is valid up to a dust-to-gas ratio of 0.15 in our case. If this is a real transition remains to be seen in experiments approaching higher dust-to-gas ratios from the lower end.}

\section{Discussion}

In this work, we use the laboratory setup of a rotating experiment to study particle clouds at varying but in general high dust-to-gas ratios for a longer time in spite of sedimentation under Earth's gravity.
This allows an analysis of the particle motion in the presence of other particles for different conditions. 
The basic parameter that is measured is the sedimentation velocity of all individual particles and their distances (in 2D).
Variations in the \textit{local} dust-to-gas ratio are important, but this value is not sufficient to describe the motion of a local particle cloud as it neglects the influence of neighboring regions on the particle motion. We therefore define the closeness as another parameter for each individual particle, which depends on all particles and their distances 
to the particle considered.

We can separate two regimes of particle motion.
For dust-to-gas ratios below $\epsilon = 0.08$ particles behave like test particles embedded in a gas. That means that they essentially do not feel the presence of the other particles, or in other words they sediment like individual particles would sediment. There is no collective effect. This even holds for regions of closeness up to 20 $\rm mm^{-1}$. 
As closeness might not be an intuitive quantity, it might be argued that the highest closeness might still be representative of a dilute cloud corresponding to test particle behavior. However, this argument does not hold for denser clouds.

For dust-to-gas ratios above 0.08 {(see fig. \ref{fig.sensitive1})} all particle motion does depend on closeness. So for
high dust loading, particles of closeness well below 20 $\rm mm^{-1}$ also have much higher sedimentation speeds than individual test particle grains should have, and they vary with the closeness of the grain. 
To first order, the speed is now also linear {with} closeness of small values, and the {slope} of this linear dependence is sensitive to the global dust-to-gas ratio or average closeness. 

We therefore define the absolute value of the {slope} as the sensitivity factor.
The sensitivity factor becomes zero at a dust-to-gas ratio $\epsilon \le 0.08$, which quantifies the
separation of the two regimes of test particle behavior for lower values and collective
motion for higher values.

{
The influence of closeness on sedimentation velocity is reminiscent of similar findings on sedimenting clouds in the Stokes regime. There, all particles perturb their surrounding-gas velocity inversely proportionally to the distance of the gas, leading to a correction term in the sedimentation velocity on the order of $O(1/r)$ \citep{GUAZZELLI2006539}. However, in \citet{GUAZZELLI2006539}, a well-confined sedimenting cloud forms a torus and finally breaks up. We do not observe anything like this, and consider the situation to be different {in terms of the experimental setup, local particle volume filling factor and local particle density}. 

Nevertheless, as a plausibility check we might consider the following simplified picture. 
If the global dust-to-gas ratio is sufficiently high, the total cloud is sufficiently {\em opaque} to the gas and can be considered as a single particle in the Stokes regime. In contrast, at later times with a low average dust-to-gas ratio, the cloud becomes {\em geometrically thin} and gas streams through it unhindered, and the cloud can no longer act as one big Stokes particle. Therefore, in the first case (Stokes regime) the closeness plays a role, whereas in the second case (Knudsen regime) no such effect can be observed. 

As a criterion on when the cloud is "opaque" to the gas, one can estimate the velocity change $\delta u$ the gas undergoes as it streams through a particle cloud of dimension $a$ at a given dust-to-gas ratio $\epsilon$:
 }
\begin{equation}
u = - \epsilon \frac{v_0}{\tau_f} t_a.
\end{equation}
With $t_a = a / v_0$ being the crossing time for gas through the clump, we can claim that significant opaqueness is given when $u = v_0$
and thus:
\begin{equation}
\epsilon_c  = \frac{v_0 \tau_f}{a}.
\end{equation}
{Plugging in a cloud size from Fig.\ (7) with about $a=40~\rm mm$, $v_0 = 70~\rm mm s^{-1}$, and $\tau_f = 7\cdot 10^{-3}~\rm s$ results in a critical density of $\epsilon_c = 0.01$, and we see that the actual $\epsilon$ values indicate easily that this clump cannot be penetrated by the gas. 
At late times when the clumps are no longer as big as initially, say for instance at $a = 5~\rm mm$, the critical  value for dust-to-gas would be $\epsilon_c = 0.1$, which is on the order of the transition in Fig. (9), when collective effects involving closeness are no longer observable. This is only a first qualitative reasoning.
Certainly, there is, for example, a limit on increasing the size of a cloud at low dust-to-gas ratios to trigger collective behavior. Otherwise, any part of a protoplanetary disk would easily become collective.}

In particular, at times of high mass loading in the experiments, individual particles in low-closeness regions with low sedimentation velocities 
can change their motion by entering regions of high closeness passing by, speeding them up. 
However, they can also drop out of warp again into a region of lower closeness, thus slowing down. 

We see fluctuations but do not see any concentration effect yet leading to a continuous local increase of particles. 
{Whether sedimentation induces gas flow and the associated turbulence leads to local particle concentrations in a nonlinear fashion is still an open question.

With the current design,
the experiment is biased to losing subclouds of high closeness or particles with high sedimentation speeds.  The stability point of fast particles is closer to the experiment wall, and therefore the stability region is much smaller. Such regions are therefore preferentially lost due to collisions with the walls. This is also visible in the evolution of the closeness distribution, where large values decrease more strongly than the small ones, leading to a nonsymmetric distribution with a small slope rising but a steep slope falling off toward higher closeness. In a way, the system cleans itself of 
very ''close'' or ''unstable'' regions. On the other hand, the difference in centrifugal motion responsible in some way mimics streaming, as denser cloudlets move further outward and can collect individual grains}.

%The sensitivity factor dependence indicates that the average closeness might be better suited as a description for the dense regime than the average dust-to-gas ratio. At least the sensitivity factor is linear for all values of average closeness, while there is a deviation from linearity for the highest dust-to-gas ratio. 

As far as sedimentation velocity goes, the variation (width) is linear in the absolute value (fig. \ref{fig.fluctuate}). 
However, the closeness gets more pronounced at higher dust-to-gas ratios, but currently we would not give this a fundamental interpretation in view of the experimental constraints on cloud stabilities.

Boiling this down, the essence of this work is the finding of increasing sensitivity
with an increasing dust-to-gas ratio. For dilute systems, a local increase in particle density -- or better, in closeness -- has little effect. At high particle loading, small
changes in particle density lead to large changes in the sedimentation velocity of that region, and more so in denser region.

This implies that some self-amplification {of the fluctuations in $\epsilon$} could only work where denser regions move faster,
pick up grains in slower, less dense regions, and move even faster. So far we are only approaching this state, as these
regions are unstable in the experiment and grains are lost to the walls.
It is a strong indication, though, that some 
sedimentation, and streaming instabilities, e.g. with centrifugal motion, might occur in laboratory
experiments.

\section{Summary}

To sum this up, for a given experiment we can describe the absolute value of the sedimentation velocity of a particle as 
\begin{equation}
v = v_0  \quad \texttt{for} \quad \epsilon < \epsilon_{\rm crit}=0.08\\
\end{equation}
\begin{equation}
v = v_0 + \alpha \cdot (\epsilon-\epsilon_{\rm crit}) \cdot C  \quad \texttt{for} \quad 0.15 \geq \epsilon \geq \epsilon_{\rm crit}=0.08. 
\label{eq.collective}
\end{equation}
The value of $\alpha$ is about $10.3 \rm mm^2/s$.

%This might be compared to the simple assumption of a single particle clump of size $r_c$ sedimenting in an otherwise undisturbed gas.
%\begin{equation}
%v = v_0 + \beta \cdot (\epsilon - 0) \cdot r_c^2
%\end{equation}

%There are two principle differences. First, we do have a minimum solid to gas ratio, $\epsilon_{crit}$, for which collective motion gets important. Second, we traded the clump radius, $r_c$, for the closeness, $C$, as we do not have a well defined clump size, and closeness accounts for the local particle density as well as the distance between grains. 

\section{Caveats and Future Work}

There are limits in this work which might be improved in the future. 
The observations are only two-dimensional, and the particle distribution along the rotation axis is unknown. So the absolute dust-to-gas ratios might have been systematically estimated to be too high. This can be improved with an additional camera and viewing angle.

Even after much testing, this work is essentially only based on the analysis of a single experiment (except the examples of particles entering and dropping out of a dense region), as we have to develop an appropriate way of analyzing dense many-particle systems, and to see where this might be leading. {Certainly, the database has to be extended to study the different parameters mentioned. This is not meant to raise doubts that what we call a single experiment is statistically insignificant. For the studied parameters one experiment provides a large data set, and the given results on transitions to collective behavior are valid.}

There might be interesting physics in the dense (high-closeness) regions that 
are lost early on. It might be that it is exactly the regions of high closeness early on, which are lost due to instability in our experiment, that correspond to unstable regions in the sense of a sedimentation instability. This is only a speculation at this point, though, as it might be due to the initial conditions and might not have evolved into this state on its own. Dense regions might be kept longer if the experiment Stokes number is decreased, decreasing centrifugal losses. 
A decrease is possible but either requires slower rotation speeds, higher gas pressure, or smaller particles.
The experimental capabilities still have to be explored. 

Also, future experiments with low-density clouds which are compressed or enhanced afterward by some means will provide deeper insight.

We start with the premise that at low particle loading, particles move 
through a gas in rigid rotation, defining a relative sedimentation velocity. The gas motion will change in a more complex way in the case of high mass loadings. After all, it is the gas that mediates the coupling between grains. The calculated sedimentation velocities are therefore still a well-defined construction, but at some point might lose the simple interpretation of relative velocity between grains and gas. 

To make the laboratory findings and laboratory extension more applicable to protoplanetary disks, we intend to numerically simulate the experiment, gas, and particles.

{The fact that the individual particles are in the Knudsen regime, whereas the clumps are in the Stokes regime is the typical case for a protoplanetary disk. Thus studying this transition in experimental setups and accompanying numerical simulations will help us to also produce more reliable predictions of the occurrence of the streaming instability.}

{Electrostatic charging of the grains might have an influence on the cloud evolution. We cannot exclude the fact that grains are charged during injection. In fact, 
\citet{yoshimatsu2017} used similar hollow spheres and showed that particles charge up after several seconds of intense vibration.
However, they only observed effects under microgravity. Also \citet{Jungmann2018} showed charging occurs in collisions under microgravity. Charging  is certainly important during sticking events \citep{lee2015}. We do not observe trajectories, though indicating the attraction or repulsion of grains, and essentially exclude further collisional charging and sticking, {but we consider our} system to be collisionless. 

{The collision time is approximately\\ 
$( \frac{N_P}{V}\cdot (4\pi d_P^2\cdot) \Delta v_p )^{-1}$ $= 14~\rm s$ with a maximum number of particles $N_P = 650$, a volume $V = 31800$~mm$^3$, a particle diameter $d_P=650\, \rm \mu \rm m$ and a relative particle velocity of $10~\rm mm s^{-1}$}. {Unnoticed collisions might result in changes in the sedimentation velocity independent of the closeness. Such collisions would lead to only one outlying data point.  We evaluated almost one million data points and therefore consider collisions to be negligible.}  Also, we do not observe the formation of aggregates.}

\section{Conclusion}

We define a Stokes number {in the experiment} with respect to the experiment rotation timescale, which turns out to be 0.014. 
This is a different Stokes number from the one defined in disks around young stars to describe the triggering of streaming instabilities for an estimated critical dust-to-gas ratio of $\epsilon = 0.027$ \citep{yang2017}.

{Local concentrations of particles in both situations require local turbulent Stokes numbers of order unity. This means that particles get concentrated when their friction time is on the order of the local turbulent correlation time. This is the case for the so far only numerically studied streaming instability in disks around young stars. Whether the sedimentation process studied in our experiments will also lead to velocity fluctuations and correlation times on the order of the friction time still has to be studied.}

We see no collective behavior for low dust-to-gas ratios up to $\epsilon = 0.08$, which we interpret as a threshold for the gas in the Knudsen regime to 
recognize a collection of particles as an obstacle to the flow.
However, above this value {(being well within the Stokes regime for the clumps)}, all dust motion is increasingly sensitive to small disturbances of the particle's closeness, which
{may or may not lead to} unstable situations. 

Overall, our laboratory experiments are but a first study. Nevertheless, this 
is the first experimental study of the underlying common-ground physics of sedimentation and streaming instabilities at the transition from the Knudsen to the Stokes regime.

The results indicate that there is potential for future improvements, but we clearly approach collective behaviors. Further experiments will hopefully lead us to well-confirmed pictures of particle concentration and planetesimal formation by self-gravity and streaming instability.
\begin{acknowledgements}
This project is supported by DFG grant WU 321/16-1 and Kl 1469/14-1.
We thank the two referees for their thorough review of this work.
\end{acknowledgements}

\bibliographystyle{aasjournal}      % mathematics and physical sciences
%\bibliography{references}   % name your BibTeX data base

%\begin{thebibliography}{}

% etc
%\end{thebibliography}

\end{document}